\documentclass[draft]{agujournal2019}
\usepackage{url} 
\usepackage{soul}
\usepackage{amsmath}
\usepackage{bm}

\soulregister\ref7
\soulregister\cite7
\soulregister\citeA7

\draftfalse

\journalname{JGR: Planets}

\begin{document}

%
%


\title{Magmatic intrusions control Io's crustal thickness}

%
%




\authors{D. C. Spencer\affil{1}, R. F. Katz\affil{1}, and I. J. Hewitt\affil{2}}


\affiliation{1}{Department of Earth Sciences, University of Oxford, Parks Road, Oxford OX1 3PR, UK}
\affiliation{2}{Mathematical Institute, University of Oxford, Woodstock Road, Oxford OX2 6GG, UK}




\correspondingauthor{Dan Spencer}{dan.spencer@earth.ox.ac.uk}




\begin{keypoints}
\item We present a model coupling magmatic segregation, compaction, magmatic intrusions, and eruptions for Io.
\item Magmatic intrusions deliver heat to the lower crust, controlling crustal thickness.
\item Potential observations of a high-melt-fraction region can be explained as a decompacting boundary layer.
\end{keypoints}

%
%

%
%


\begin{abstract}
Io, the most volcanically active body in the solar system, loses heat through eruptions of hot lava. Heat is supplied by tidal dissipation and is thought to be transferred through the mantle by magmatic segregation, a mode of transport that sets it apart from convecting terrestrial planets. We present a model that couples magmatic transport of tidal heat to the volcanic system in the crust, in order to determine the controls on crustal thickness, magmatic intrusions, and eruption rates. We demonstrate that magmatic intrusions are a key component of Io's crustal heat balance; around 80\% of the magma delivered to the base of the crust must be emplaced and frozen as plutons to match rough estimates of crustal thickness. As magma ascends from a partially molten mantle into the crust, a decompacting boundary layer forms, which can explain possible observations of a high-melt-fraction region.
\end{abstract}

\section*{Plain Language Summary}
Io is cyclically deformed by Jupiter's gravity as it orbits. This deformation causes heating, which is the energy source for Io's spectacular volcanism. The tidal energy causes rock to melt inside Io, but how this magma is extracted from the mantle and onto the surface is not well understood. In this work we use a mathematical model to quantify a hypothesis for how magma (and the energy it carries) moves from Io's interior to the surface. We show that a large proportion of magma must be freezing in the crust. If this weren't the case, Io would lose its heat so efficiently that the crust would becomes hundreds of kilometers thick, much thicker than is considered reasonable on the basis of the abundant surface volcanism. Further, previous studies have proposed that a ``magma ocean'' exists beneath the crust. We show that high magma pressure causes magma to accumulate into a magma-rich layer, providing a possible explanation for how this layer of high melt fraction formed.

%
%

%


%
%
%
%

\section{Introduction}
Jupiter's moon Io is tidally heated by eccentricity forcing from its mean motion resonance with Europa and Ganymede \cite{lainey_strong_2009}, resulting in extensive surface volcanism. This volcanism has lead to significant interest in understanding Io's internal structure and energy balance. The rate of tidal dissipation coincides closely with the rate of surface heat loss \cite{davies_map_2015}, implying that Io is close to a state of thermal equilibrium. Further, the surface is crater free with globally distributed, low-relief volcanoes, implying relatively uniform global resurfacing. These observations imply that Io's leading-order structure is spherically symmetric and roughly steady state. An understanding of this leading-order structure must serve as the foundation for investigations into spatial heterogeneity and temporal evolution.

Io's radial structure is determined by the heat and mass transport mechanisms operating in its interior. Energy emission from the surface is concentrated at volcanic features, suggesting that volcanism, not conduction, is the primary heat transport mechanism in the crust. \citeA{oreilly_magma_1981} proposed that the export of heat across the crust by volcanic systems --- a process known as heat piping --- allows the growth of a thick crust, which limits the efficiency of conductive heat loss.

Heat transport in Io's mantle is more widely debated. A thermal equilibrium requires that the rate of energy export matches the rate of tidal dissipation. \citeA{moore_tidal_2003} demonstrated that an equilibrium between convective heat transport and tidal dissipation would occur at melt fractions above disaggregation. However, the expected tidal heat production under these conditions is significantly less than the observed surface heat flux. This suggests that convection cannot be the primary mechanism for delivering heat to the crust. Alternatively, magmatic segregation is capable of transporting the observed tidal heat input at low melt fractions \cite{moore_thermal_2001,breuer_10.08_2015}. It is therefore plausible that Io's tidal heat is removed from the mantle by magmatic segregation and transported across the crust by a volcanic plumbing system. We adopt this basic hypothesis in the current study and specifically address what controls the total amount of magma produced, the amounts emplaced intrusively as plutons and extrusively as surface volcanism, and the controls these place on crustal thickness.

The thickness of Io's crust is determined by the depth to which it downwells before it is heated to its melting point. Previous work has not considered the dynamics of magma at the crust--mantle boundary or in the lower crust, but the emplacement of plutons introduces heat to the lower crust, which reduces the crust's thickness and modifies its thermal profile. We present a coupled model of crust and mantle dynamics that assesses these processes. The model is formulated to make predictions of elastic thickness, surface heat fluxes, and globally averaged eruption rates, predictions that can be readily tested by future missions to the Jupiter system. Our results indicate that the heat balance and melt-transport mechanisms in the crust and at the crust--mantle boundary ultimately determine the thickness of Io's crust and the melt distribution below it.

Magnetic induction measurements --- interpreted in terms of mantle electrical conductivity --- have been used to infer the presence of a layer around 50~km thick with more than $20$\% melt fraction (a ``magma ocean'') beneath Io's crust \cite{khurana_evidence_2011}. We note, however, that the interpretation of the induction measurements as a high-melt-fraction region is debated; \citeA{blocker_mhd_2018} argue that interaction with Io's plasma environment is a better explanation of induction measurements than is a magma ocean. Nonetheless, previous studies have proposed that this inferred high-melt-fraction layer could be a region of enhanced tidal dissipation \cite{hamilton_spatial_2013,bierson_test_2016}. Tidal dissipation theory predicts that for a homogeneous body, dissipation is highest at the center. A low viscosity layer underlying a rigid crust may allow the concentration of dissipation, but models that invoke it must explain how such a structure arises. We use our coupled dynamical model of the crust and mantle to investigate the feasibility of a high-melt-fraction layer occurring without the need for enhanced dissipation.

The manuscript is organised as follows. First we outline the physics of the model before presenting results showing the key controls on \textit{i}) crustal thickness, \textit{ii}) emplacement rates, and \textit{iii}) a high-melt-fraction layer beneath the crust. We then discuss the implications of these results for interior structure and evolution.

\section{Model Description}
\begin{figure}
\centering
\includegraphics[scale=0.15]{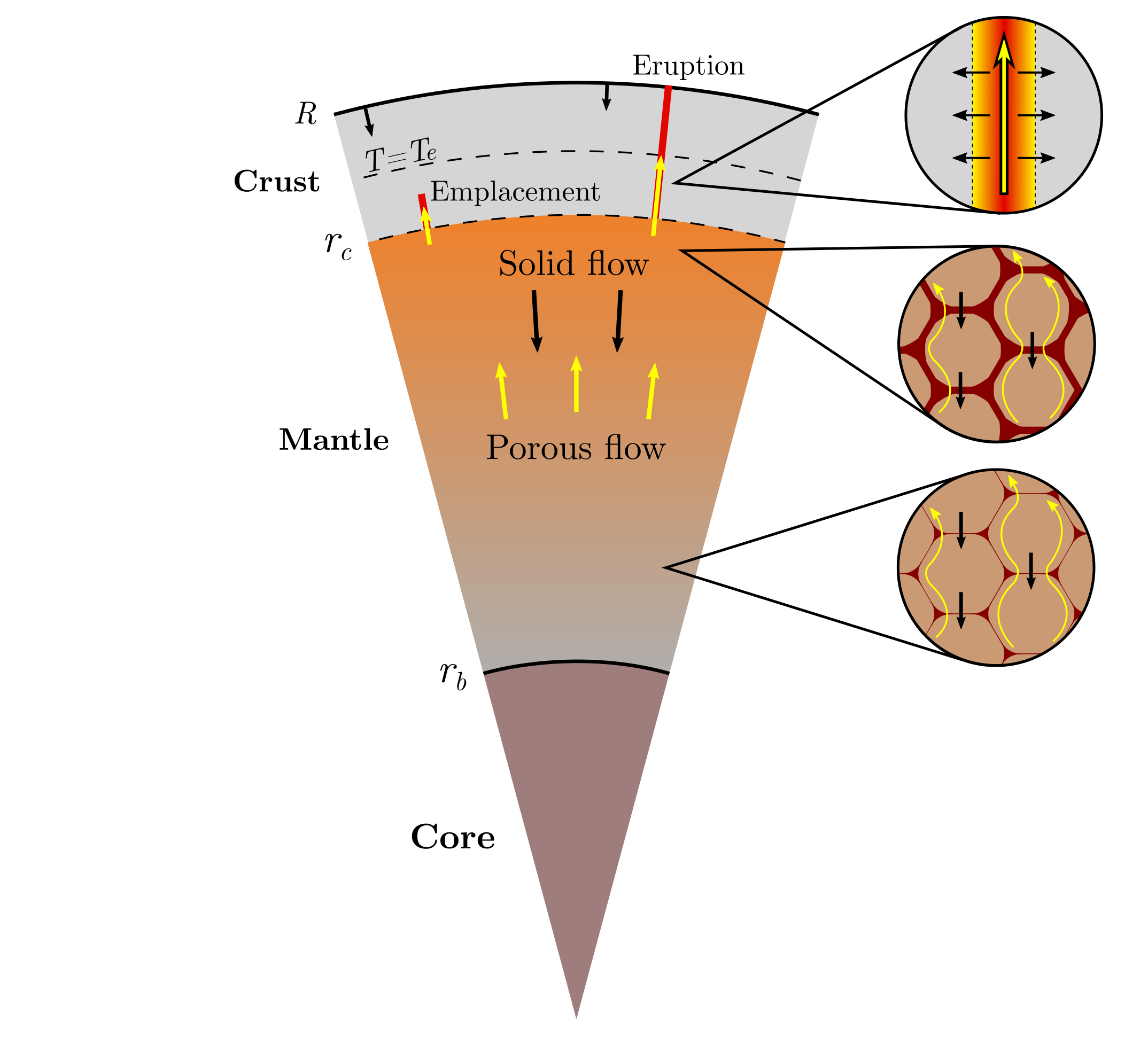}
\caption{Schematic of the model for Io. Magma rises buoyantly through the mantle while the solid moves downward. At the top of the mantle, magma enters the volcanic plumbing system. Some of this magma is emplaced (intruded) into the ductile lower crust; the rest rises to the surface and fuels volcanic eruptions. The core is excluded from the model.}
\label{schematic}
\end{figure}

The model, shown schematically in figure \ref{schematic}, considers the dynamical effects of tidal dissipation on Io's crust and mantle. These are modelled as a continuum that is either solid (the crust) or partially molten (the mantle). We model melting, magmatic segregation, and compaction (the contraction of a solid matrix as melt is expelled) with a system of conservation equations for mass, momentum, and energy appropriate for a compacting two-phase medium \cite{mckenzie_generation_1984}. Magmatic flow can also occur in a volcanic plumbing system that stretches from the upper mantle to the surface, and that exchanges mass with the mantle and crust. In the upper mantle, melt can leave the pore space and enter the plumbing system and then, as melt rises in the plumbing system, it can form intrusions (freeze) in the crust, delivering mass and energy to the surroundings. The volcanic flux that reaches the surface (the eruptive flux) instantly cools and imparts a downward flux of cold surface material. The crust--mantle boundary is defined as the depth at which the temperature is equal to the solidus temperature (see table \ref{table:parameters1}); its location is determined as part of the model. Crustal thickness is thus defined as the distance over which the cold, downwelling surface material heats before it begins to re-melt. This is not a typical definition of crustal thickness, but it seems the most natural in the present context of a one-component thermochemical model, which precludes a petrological definition.

The model invokes some simplifying assumptions. Io's volcanoes are distributed across its entire surface \cite{kirchoff_global_2011,williams_volcanism_2011}; this and the lack of craters implies global resurfacing. Though it has been noted that there is a degree-2 pattern to hotspot locations \cite{kirchoff_global_2011,hamilton_spatial_2013,rathbun_global_2018}, we investigate a spherically symmetric model for consistency with Io's apparent global resurfacing. We assume that deviations from spherical symmetry are secondary effects imprinted on a leading-order radial structure. These deviations from spherical symmetry are expected to be important in discerning the global distribution of tidal dissipation\cite{veeder_io:_2012,de_kleer_time_2016,cantrall_variability_2018,rathbun_global_2018}, but are beyond the scope of this work. Tidal dissipation models that match Io's surface heat flux utilise either very low viscosities \cite{steinke_tidally_2020} or empirically parameterised rheologies \cite{bierson_test_2016,renaud_increased_2018}. To explore the leading-order dynamics without a dependence on poorly constrained parameters, we take tidal dissipation to be uniformly distributed. Below we assess the melt configurations this produces and discuss whether this may lead to significant radial partitioning of tidal heating. In order to keep the model simple, we assume one-component thermodynamics, so the composition of the rock is neglected. Future iterations of the model will incorporate multiple components and allow us to investigate segregation between components. We neglect the pressure-dependence of the melting temperature due to the small size of Io and hence the low pressures in the mantle. Finally we assume that melt is mobile in the partially molten mantle due to the large grain size, and hence large permeability, expected for a refractory, annealed mantle \cite{lichtenberg_magma_2019}.

In the mantle (radii $r_{b}<r<r_{c}$), temperature is at the melting point and heat transport occurs solely by magmatic transport of latent heat. Buoyancy causes the upward flow of magma, which is balanced by the downward flow of solid. In the crust ($r_{c}<r<R$, defined as having a spherically-averaged temperature below the melting point), the temperature drop to the surface drives a conductive heat flux, while the downwelling solid crust transports the cold surface temperature inward \cite{schenk_origin_1998}. The volcanic plumbing system continues to transport magma and latent heat upward through the crust. We make minimal assumptions about what this plumbing system actually looks like, but require that it interacts with the solid crust by emplacement of material (the formation of plutonic intrusions). This emplaced material is a crustal source of both mass and heat. The actual volume of the plumbing system is assumed to be negligible (consistent with the flow there being much faster than that of the solid and melt elsewhere).

\subsection{Model equations}
In the partially molten mantle, which has melt fraction $\phi(r,t)$, we represent the solid velocity by $\bm{u}$ and the relative liquid velocity (the Darcy segregation flux) by $\bm{q}=\phi(\bm{v}_{\textrm{liquid}}-\bm{u})$. Darcy's law relates the segregation flux to pressure gradients and buoyancy
\begin{linenomath*}
\begin{equation}
    \bm{q} = -\frac{K_{0}\phi^{n}}{\eta_{l}}\Big[(1-\phi)\Delta \rho \bm{g} + \bm{\nabla} P \Big],
    \label{eq:Darcy}
\end{equation}
\end{linenomath*}
where $K_{0}\phi^{n}$ is the permeability, $n$ is the permeability exponent, $\Delta \rho$ is the density difference between solid and liquid, $\bm{g}=-g\hat{\bm{r}}$ is the gravity vector, $\eta_{l}$ is the liquid viscosity, and $P=(1-\phi)(P_{\textrm{liquid}}-P_{\textrm{solid}})$ is the compaction pressure \cite{keller_numerical_2013}.

Transfer of material between the crust--mantle system and the volcanic plumbing system is considered in the conservation of mass equation. Making a standard Boussinesq approximation (that is, ignoring the density difference except were it appears in the body force term of equation (\ref{eq:Darcy})), conservation of solid and liquid mass require
\begin{linenomath*}
\begin{equation}
    \frac{\partial}{\partial t}(1-\phi) + \bm{\nabla}\cdot[(1-\phi)\bm{u}] = -\Gamma + M,
    \label{eq:mass_s}
\end{equation}
\end{linenomath*}
\begin{linenomath*}
\begin{equation}
    \frac{\partial\phi}{\partial t} + \bm{\nabla}\cdot (\phi \bm{u}+\bm{q}) = \Gamma - E,
    \label{eq:mass_l}
\end{equation}
\end{linenomath*}
where $\Gamma$ is the volume transfer rate of solid into liquid (the melting rate), $E$ is the extraction rate to the plumbing system (a sink of liquid from the mantle), and $M$ is the emplacement rate from the plumbing system (a source of solid to the crust). Adding equations (\ref{eq:mass_s}) and (\ref{eq:mass_l}) gives conservation of mass in the crust--mantle system
\begin{linenomath*}
\begin{equation}
	\bm{\nabla}\cdot (\bm{u}+\bm{q}) = M-E,
	\label{eq:mass_cont}
\end{equation}
\end{linenomath*}
where we have assumed that the total fraction of Io occupied by the plumbing system is negligible. The flux of material in the plumbing system $\bm{q}_{p}$ increases when material is extracted from the mantle and decreases when material is emplaced back into the crust. Conservation of mass in the plumbing system is therefore given by
\begin{linenomath*}
\begin{equation}
	\bm{\nabla}\cdot \bm{q}_{p} = E -M.
	\label{eq:consv_mass_plum}
\end{equation}
\end{linenomath*}
We assume that the magma in the plumbing system is at the melting point $T_{m}$ and we parameterise the emplacement rate based on the temperature difference to the host material at temperature $T$,
\begin{linenomath*}
\begin{equation}
	M = \begin{cases} \frac{hc(T_{m}-T)}{L} & T>T_{e}, \\ 0 & T<T_{e}, \end{cases}
	\label{eq:emplacement}
\end{equation}
\end{linenomath*}
where $c$ is the specific heat capacity, $L$ is the latent heat, $h$ is an emplacement rate constant (units s$^{-1}$) and $T_{e}$ is the elastic-limit temperature. Regions of the crust colder than the elastic-limit temperature are assumed to be brittle; magma propagates through these without permanent emplacement. This assumption is introduced to highlight and investigate the importance of the distribution of magmatic emplacement in controlling the crustal temperature profile. The detailed mechanisms of dike propagation and emplacement are subsumed in this parametrisation. Although $h$ could plausibly be related to heat conduction in the vicinity of a dike, we treat it here as a free parameter and explore the model's behaviour for a wide range of values.

Extraction of liquid from the mantle into the plumbing system is expected to take place at the top of the mantle. We assume that this transfer is a function of liquid overpressure,
\begin{linenomath*}
\begin{equation}
	E = \begin{cases} \nu (P-P_{c}) & P>P_{c}, \\ 0 & P<P_{c}, \end{cases}
\end{equation}
\end{linenomath*}
where $\nu$ is an extraction rate constant (units s$^{-1}$Pa$^{-1}$) and $P_{c}$ is a critical overpressure that liquid must attain in order to be extracted into the plumbing system. Liquid overpressure (compaction pressure) is related to the compaction rate $\bm{\nabla}\cdot \bm{u}$ by the relationship \cite{mckenzie_generation_1984}
\begin{linenomath*}
\begin{equation}
    P = \zeta \bm{\nabla}\cdot \bm{u},
    \label{eq:pressure_relation}
\end{equation}
\end{linenomath*}
where $\zeta = \eta/\phi$ is the compaction viscosity, related to the shear viscosity $\eta$. This form of the compaction viscosity is commonly assumed, but other forms have been proposed with a weaker singularity as $\phi \rightarrow 0$ \cite{rudge_viscosities_2018}.

We model heat transport in the crust and mantle of Io together, using an enthalpy method \cite{katz_magma_2008} so that no boundary conditions need be imposed an the crust--mantle boundary. Conservation of energy requires
\begin{linenomath*}
\begin{equation}
    \frac{\partial H}{\partial t} + \bm{\nabla}\cdot [(\bm{u}+\bm{q})T] + \frac{L}{c}\bm{\nabla}\cdot \left(\phi \bm{u} + \bm{q}\right) = \bm{\nabla}\cdot (\kappa \bm{\nabla} T) + \frac{\psi}{\rho c} - E\left(T+\frac{L}{c}\right) + M\left(T_{m}+\frac{L}{c}\right),
    \label{eq:energy_cont}
\end{equation}
\end{linenomath*}
where bulk enthalpy is defined as $H = T + L\phi/c$, and $\kappa$ is the thermal diffusivity. Changes in bulk enthalpy are caused by advection of sensible heat, advection of latent heat, diffusion, tidal heating, extraction of melt, and emplacement of melt, which are represented respectively by each term in equation (\ref{eq:energy_cont}). The integral of the tidal heating rate $\psi$ over silicate Io gives the total tidal heating input $\Psi$, which we take to be $1\times 10^{14}~$W \cite{lainey_strong_2009}.

\subsection{Solution methods}
The full model to be solved comprises the enthalpy equation (\ref{eq:energy_cont}) for $H$, the combination of equations (\ref{eq:Darcy}) and (\ref{eq:pressure_relation}) in a compaction equation for $P$ and $\bm{q}$, total mass conservation (\ref{eq:mass_cont}) for $\bm{u}$, and conservation of mass in the plumbing system (\ref{eq:consv_mass_plum}) for $\bm{q}_{p}$. The enthalpy solution gives $\phi$ and $T$ through the definition of bulk enthalpy $H=T+L\phi/c$, by assuming that temperature is buffered to the melting point when enthalpy exceeds that of the melting temperature; melt fraction is zero wherever temperature falls below the melting point. Parameter values used in the model are given in table \ref{table:parameters1}. The system is scaled (see appendix A) and solved using the Portable, Extensible Toolkit for Scientific Computation (PETSc) \cite{petsc-user-ref,petsc-web-page,petsc-efficient}. Robust convergence is obtained by splitting the system of governing equations into three non-linear problems for enthalpy, pressure, and plumbing-system flux. These are solved iteratively at each timestep until the L$_{2}$-norm of their residual vectors are all below a small tolerance ($10^{-7}$). We run the model to steady state and report the final, steady solutions. To facilitate exploration of the parameter space, an asymptotic approximation of these solutions is also employed. This is developed in appendix B.

\begin{table}[ht]
\caption{Dimensional parameters}
\centering
\begin{tabular}{l l l l l}
\hline
Quantity & Symbol & Definition & Preferred Value & Units \\
\hline
Radial position & $ r $ & & & m \\
Radius & $R$ & & $1820$ & km \\
Core radius$^{1}$ & $r_{b}$ & & $700$ & km \\
Crustal radius & $r_{c}$ & & & m \\
Boundary layer coordinate & $ Z $ & & & m \\
Solid velocity & $ u $ & & & m/s \\
Segregation flux & $ q $ & $ q = \phi(v_{\textrm{liquid}}-u) $ & & m/s \\
Volcanic plumbing flux & $ q_{p} $ & & & m/s \\
Porosity & $ \phi $ & & & \\
Permeability constant$^{2}$ & $ K_{0} $ & $ K = K_{0}\phi^{n}$ & $10^{-7}$&m$^{2}$ \\
Permeability exponent$^{2}$ & $n$ & see above & 3 & \\
Density & $\rho $ & & $3000$ & kg/m$^{3}$ \\
Density difference & $\Delta \rho $ & & $500$ & kg/m$^{3}$ \\
Gravitational acceleration & $g$ & & $1.5$ & m/s$^{2}$ \\
Shear viscosity & $ \eta $ & & $ 1\times 10^{20} $ & Pas \\ 
Liquid viscosity & $\eta_{l}$ & & $1$& Pas \\
Volume transfer rate & $\Gamma$ & & & s$^{-1}$ \\
Emplacement rate & $M$ & $ M = hc(T_{m}-T)/L$ & & s$^{-1}$ \\
Emplacement constant & $h$ & see above & $7$ & Myr$^{-1}$ \\
Extraction rate & $E$ & $ E = \nu(P-P_{c}) $ & & s$^{-1}$ \\
Extraction constant & $\nu$ & see above  & $1.4\times 10^{-5}$ & Myr$^{-1}$Pa$^{-1}$ \\
Compaction pressure & $P$ & $ P = \zeta \bm{\nabla}\cdot \bm{u} $ & & MPa  \\
Critical overpressure & $P_{c}$ & & $5$ & MPa \\
Compaction viscosity & $\zeta$ & $\zeta = \eta/\phi$ & & Pas \\
Temperature & $T$ & & & K \\
Elastic limit temperature & $T_{e}$ & & $1000$ & K \\
Melting temperature & $T_{m}$ & & $1500$ & K \\
Surface temperature & $T_{s}$ & & $150$ & K \\
Latent heat & $L$ & & $4\times 10^{5}$ & J/Kg \\
Specific heat capacity & $c$ & & $1200$ & J/Kg/K \\
Total tidal heating$^{3}$ & $\Psi$ & & $1\times 10^{14}$ & W \\
Tidal heating rate$^{*}$ & $\psi$ & & $4.2\times 10^{-6}$ & W/m$^{-3}$ \\
\hline
\multicolumn{5}{l}{$^{1}$\citeA{bierson_test_2016}, $^{2}$\citeA{katz_magma_2008}, $^{3}$\citeA{lainey_strong_2009}} \\
\multicolumn{5}{l}{$^{*}$ $\Psi$ divided by the volume of silicate Io}
\end{tabular}
\label{table:parameters1}
\end{table}

\section{Results}
\begin{figure}
\centering
\includegraphics[width=\linewidth]{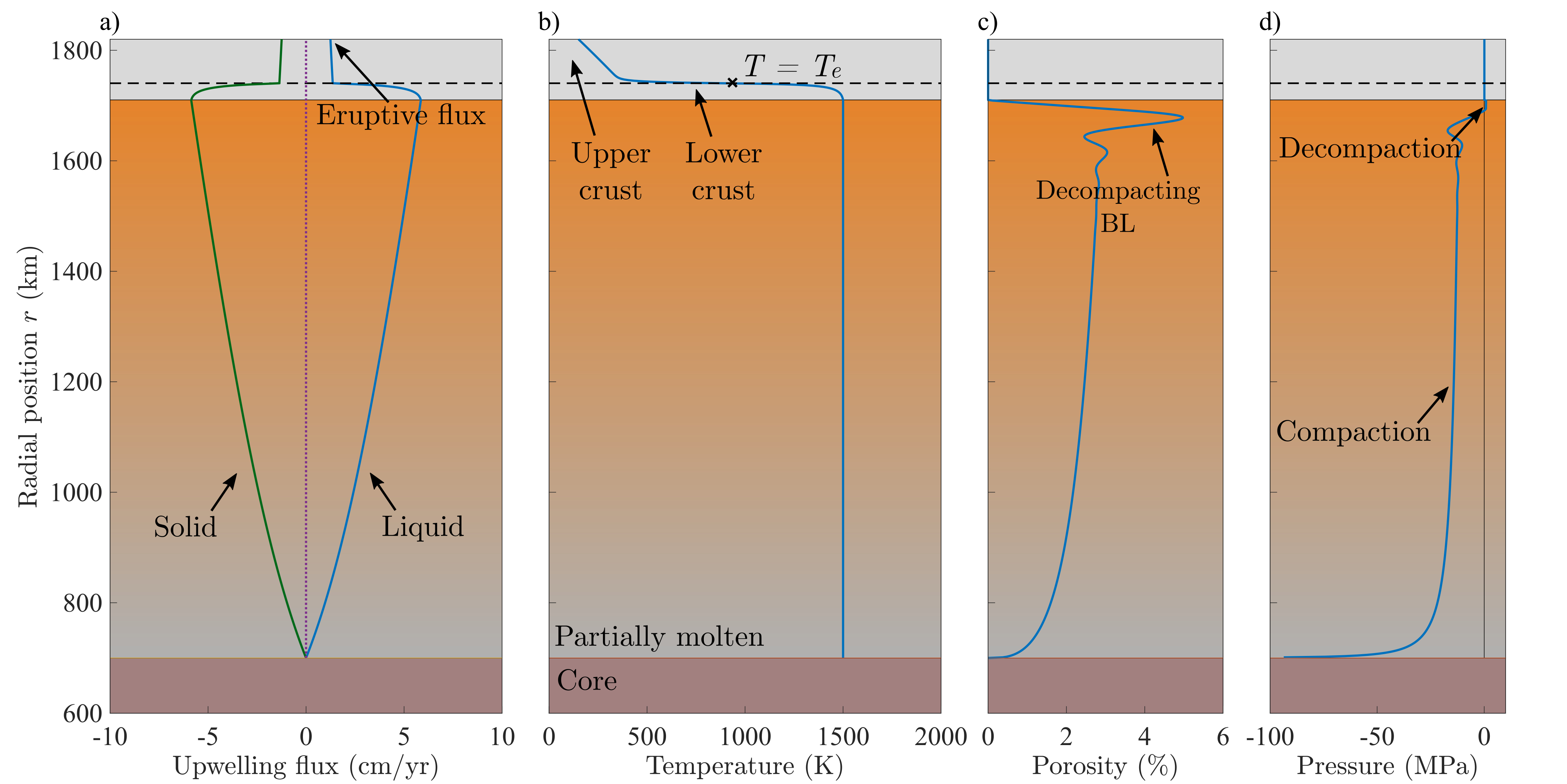}
\caption{Example solution to the model with $h=7~\textrm{Myr}^{-1}$, $T_{e}=960~$K. a) Upwelling magma is replaced by downwelling solid. b) The lower crust ($T>T_{e}$) is heated by magmatic intrusions (emplacement), but the upper crust ($T<T_{e}$) is cold due to the downwelling cold surface material. c) Melt fractions are low throughout the mantle but increase in a thin boundary layer on the order of $50$~km thick. d) Compaction occurs throughout most of mantle as the liquid is at low pressure, but beneath the crust $P\simeq P_{c}=0.8~$MPa, and downwelling solid is decompacted by liquid pressure. The elastic thickness is $80~$km and the eruption rate is $1.1~$cm/yr, with $99.5\%$ of heat transport through the surface being volcanic. Parameter values can be found in table \ref{table:parameters1}.}
\label{full_example_figure}
\end{figure}

A representative solution of the model is plotted in figure \ref{full_example_figure}. Panel (a) shows that melt upwells throughout the mantle and the solid correspondingly downwells. This process, driven by magmatic buoyancy, results in relatively low melt fractions ($\sim3\%$) except in a thin boundary layer beneath the crust that we discuss below. Magma that reaches the surface solidifies and cools to the surface temperature (with the heat released to space). The continual eruption and burial of the surface causes the crust to downwell, balancing the upward flux of magma in the plumbing system. The downwelling crust advects the cold surface temperature into the interior, resulting in a relatively cold upper crust, capable of supporting Io's mountains. As the crustal material continues to downwell through the lower crust, it is heated by magmatic emplacement (the formation of plutonic intrusions) and eventually reaches the solidus, where it starts to melt. The balance between downward advection of the cold surface temperature, and intrusive heating, results in a steady crustal thickness being maintained. A steep temperature gradient arises in the crust (fig.~\ref{full_example_figure}b) between the upper crust, where heat transport is dominated by the downward advection of cold crust, and the lower crust, where emplacement causes significant heating. Throughout most of the mantle, liquid pressure is low, causing the solid matrix to compact (fig.~\ref{full_example_figure}d). Approaching the crustal boundary, liquid pressure increases and causes the decompaction of the downwelling crustal material.

We now explore the behaviour of the model in relation to various parameters. We first investigate the effect of the parameters associated with the volcanic plumbing system, before turning to material and rheological parameters.

\subsection{Dependence on volcanic plumbing system parameters}
\begin{figure}
    \centering
    \includegraphics[width=\linewidth]{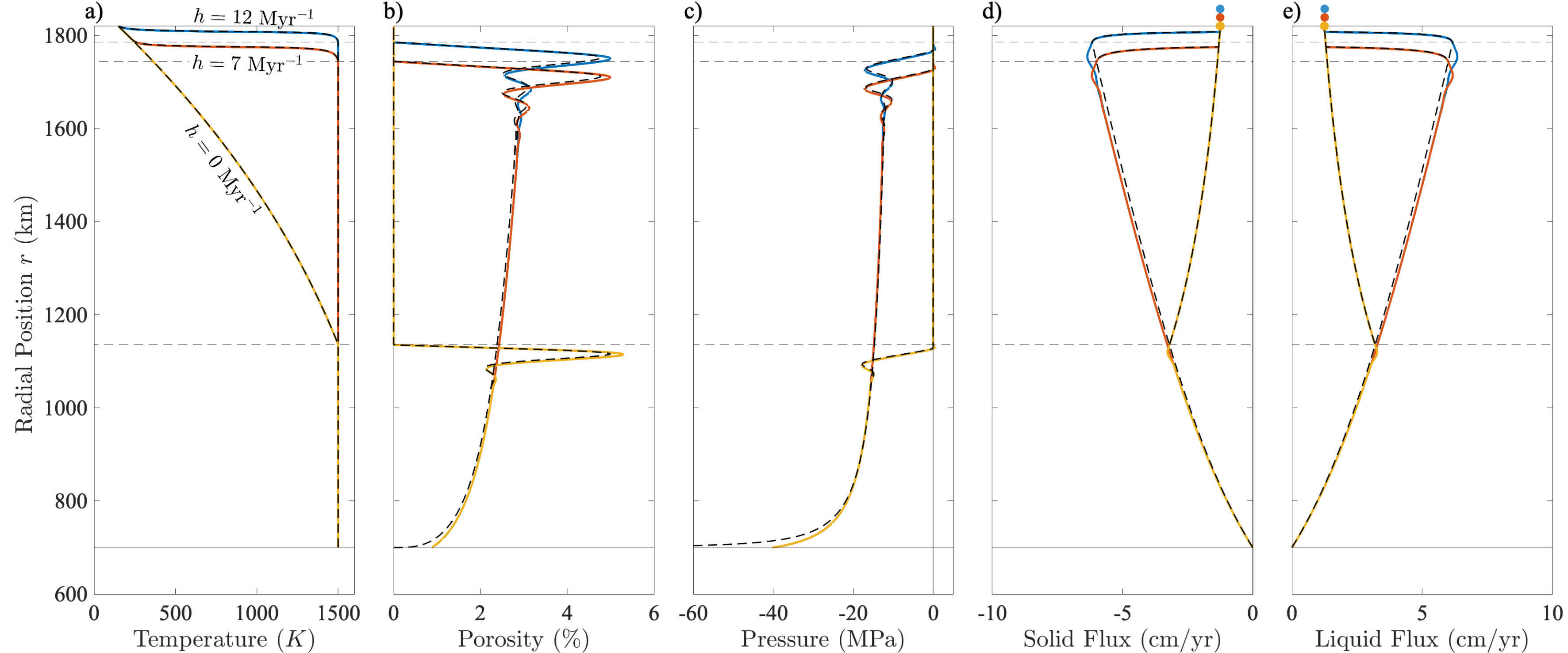}
    \caption{The effect of emplacement rate $h$ on a) temperature profile, b) porosity distribution, c) compaction pressure, d) solid and e) liquid fluxes. Solid lines are full solutions to the model, and heavy dashed lines are the approximate solutions. Thin dashed lines mark the crust--mantle boundaries. Dots on panels d and e show the surface erupted fluxes. When there is no magmatic emplacement ($h=0$), the crust grows to be over $600$~km thick due to the rapid downwelling of the cold surface temperature. As emplacement rate is increased, the heating that this provides to the crust can increasingly balance the cold downwelling surface temperature, resulting in smaller crustal thicknesses. $T_{e}=960~$K in these solutions. Liquid flux is the sum of the segregation flux $q$ and plumbing-system flux $q_{p}$. Parameters values can be found in table \ref{table:parameters1}.}
    \label{hhat_vary}
\end{figure}

\begin{figure}
    \centering
    \includegraphics[width=\linewidth]{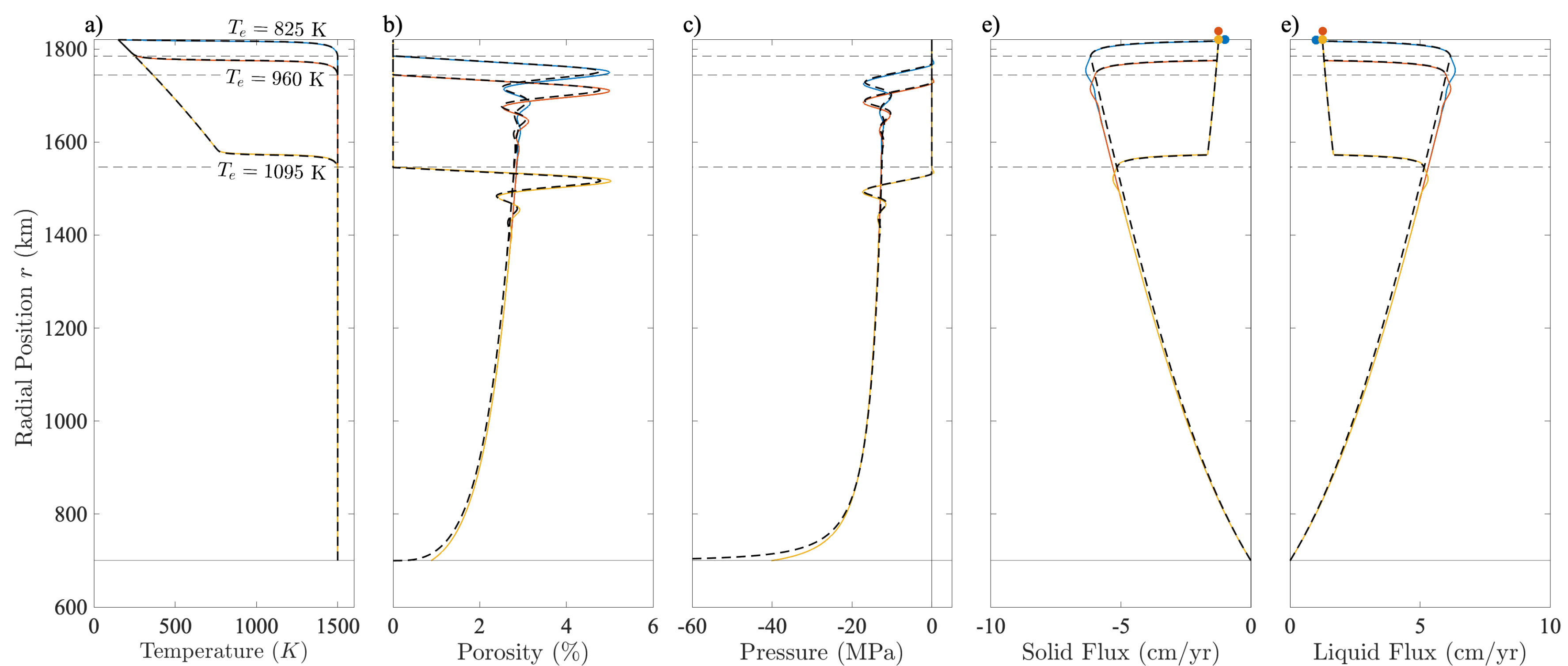}
    \caption{The effect of elastic limit temperature $T_{e}$ on a) temperature profile, b) porosity distribution, c) compaction pressure, d) solid and e) liquid fluxes. Solid lines are full solutions to the model, and heavy dashed lines are the approximate solutions. As the elastic limit temperature $T_{e}$ is increased, crustal thickness increases and a larger proportion of the upper crust is cold, resulting in larger elastic thicknesses. $h=7~$Myr$^{-1}$ in these solutions.}
    \label{Te_vary}
\end{figure}

The main parameters that control the behaviour of the volcanic plumbing system are the emplacement rate constant $h$, and the elastic limit temperature $T_{e}$. The behaviour of the system for three values of emplacement rate constant $h$ and elastic-limit temperature $T_{e}$ is shown in figures \ref{hhat_vary}--\ref{Te_vary}. The solid lines are the full, numerical solutions to the model and the dashed lines are from the asymptotic approximations (see appendix B).

Figure \ref{hhat_vary} demonstrates that the rate of magmatic emplacement $h$ exerts a strong control on Io's crustal thickness. When the emplacement rate is zero (fig.~\ref{hhat_vary}, $h=0$), magma does not re-heat the sinking crust and therefore the cold surface material downwells far into the mantle before it is heated to its melting point (by basal conduction and tidal heating only), producing a $>600~$km thick crust. Conversely, when emplacement is rapid, the crust is very quickly heated to its melting point and therefore is thin. Figure \ref{hhat_vary}e shows how the volcanic plumbing flux changes through the crust. When there is no emplacement, the drop in volcanic plumbing flux is purely due to radial spreading.

Figure \ref{Te_vary} shows the effect of increasing the elastic-limit temperature $T_{e}$. When $T_{e}$ is low, emplacement can take place throughout a large portion of the crust. This causes most of the crust to be hot (fig~\ref{Te_vary}, $T_{e}=825~$K) and thus the elastic thickness is small. Increasing $T_{e}$ leads to the growth of a large, cold upper crust where no emplacement is taking place. This provides a large elastic thickness capable of supporting Io's mountains. Figure \ref{Te_vary}e shows that the total amount of material being emplaced at steady-state does not significantly change as $T_{e}$ is increased, and the thickness of the emplacement region ($T>T_{e}$) is roughly constant as $T_{e}$ increases. As discussed in section \ref{magmagtic_intrusion}, the total amount of erupted and emplaced material is controlled by a global energy balance.

\begin{figure}
\centering
\includegraphics[width=\linewidth]{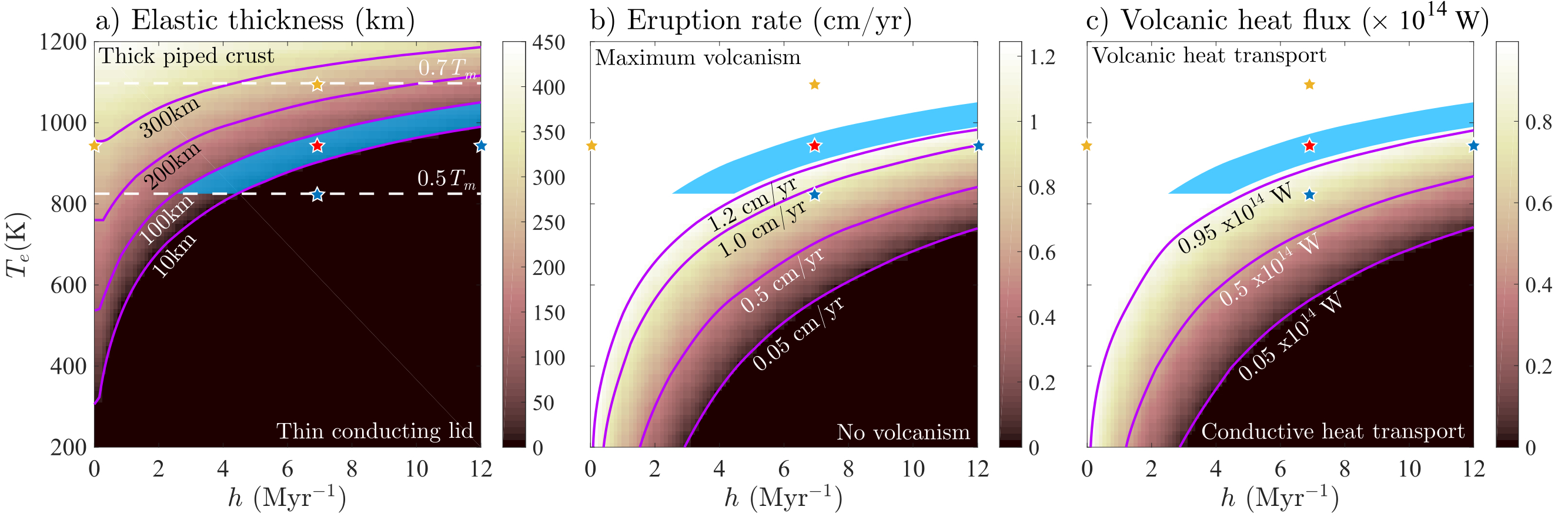}
\caption{Model predictions of a) elastic thickness, b) eruption rate, and c) volcanic heat flux in the parameter space of emplacement constant $h$ and elastic limit temperature $T_{e}$. The dashed lines on panel a) mark homologous temperatures $0.5 ~T_{m}$ and $0.7 ~T_{m}$, an estimated range for the transition from brittle to ductile behaviour. The blue region covers the parameter space for potential elastic thicknesses ($10-100$~km). The central red star marks the solution displayed in figure \ref{full_example_figure}, and other stars mark the solutions in figures \ref{hhat_vary} and \ref{Te_vary}.}
\label{h-te-plots}
\end{figure}

To comprehensively map the parameter space of $T_{e}$ and $h$ we rely on our asymptotic approximation of the steady solution. Figures \ref{hhat_vary}--\ref{Te_vary} show that there is good agreement between the full solutions and the asymptotic approximation. Figure \ref{h-te-plots} shows how (a) elastic thickness, (b) eruption rate, and (c) volcanic heat flux vary as a function of the emplacement rate constant $h$ and elastic-limit temperature $T_{e}$. Figure \ref{h-te-plots} confirms the trends seen in figures \ref{hhat_vary} and \ref{Te_vary}, and places them in the context of observable features. The blue region in figure \ref{h-te-plots} indicates the parameter space that gives reasonable elastic thicknesses ($10-100~$km) at reasonable brittle--ductile transition temperatures (homologous temperature $0.5-0.7~$T$_{m}$). Figure \ref{h-te-plots}a shows that the elastic thickness varies rapidly with relatively small changes in $T_{e}$ and $h$ around the main solution (the central star, plotted in fig.~\ref{full_example_figure}). Elastic thickness is thus the most useful observation for constraining the characteristics of Io's volcanic plumbing system. Figure \ref{h-te-plots}b shows that eruption rate reaches a maximum of $\sim 1.25~$cm/yr over much of the parameter space, a value that is discussed further below. The conductive heat flux in this part of the parameter space is negligible (fig.~\ref{h-te-plots}c) as virtually all of the input tidal heating is lost in eruptions. Figure \ref{h-te-plots}b shows that if emplacement rate is very high and takes place through the majority of the crust, eruption rate goes to zero. All heat is lost by conduction through a thin lid in this case.

\subsection{Dependence on material and rheological parameters}
\begin{figure}
    \centering
    \includegraphics[width=\linewidth]{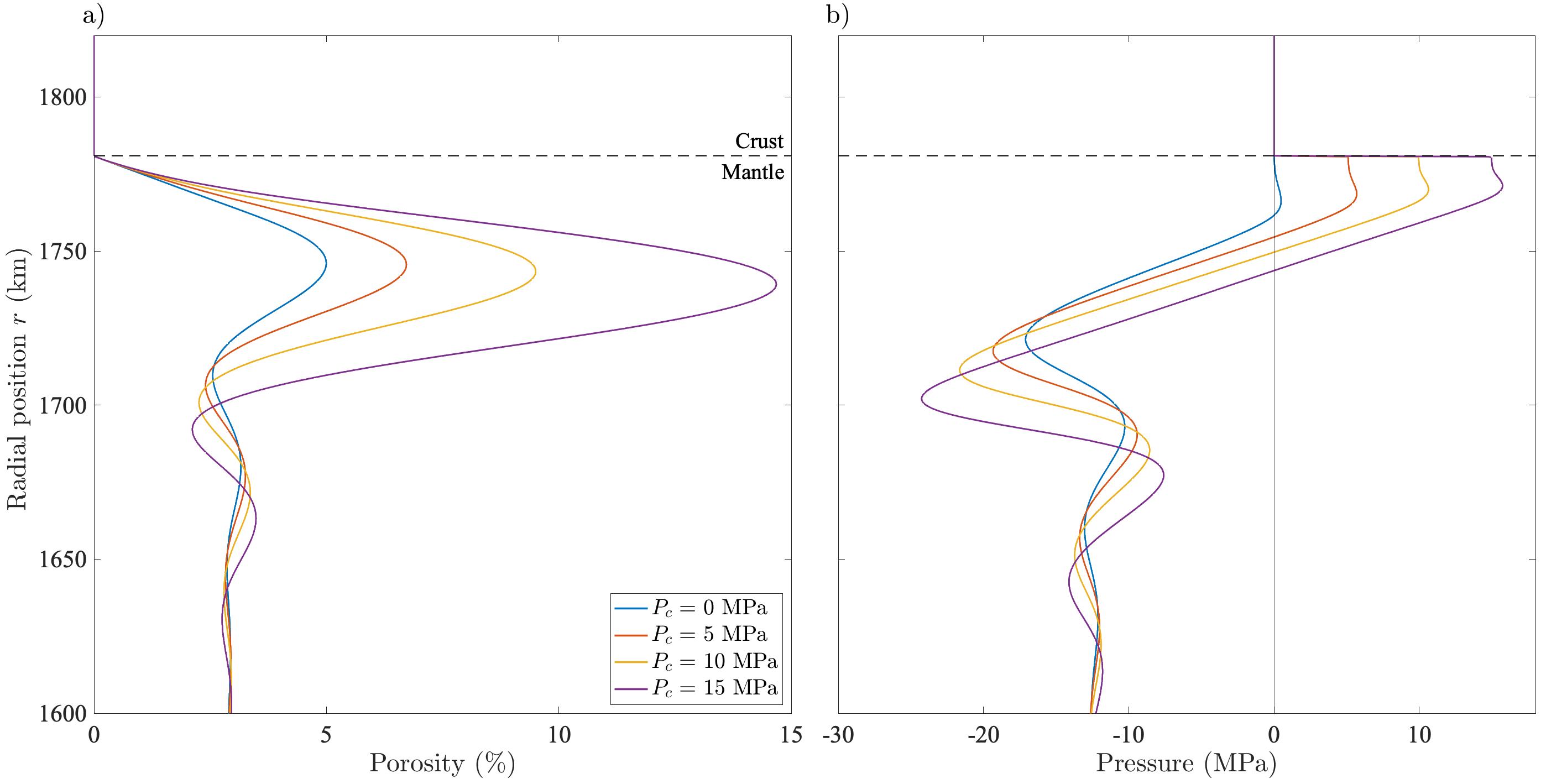}
    \caption{a) Porosity and b) compaction pressure in the decompacting boundary layer beneath the crust for different values of critical compaction pressure $P_{c}$. High values of $P_{c}$ cause the material downwelling from the crust to decompact rapidly, leading to the accumulation of large amounts of melt. Note the different radial scale from other figures.}
    \label{Pc_vary}
\end{figure}

\begin{figure}
    \centering
    \includegraphics[width=\linewidth]{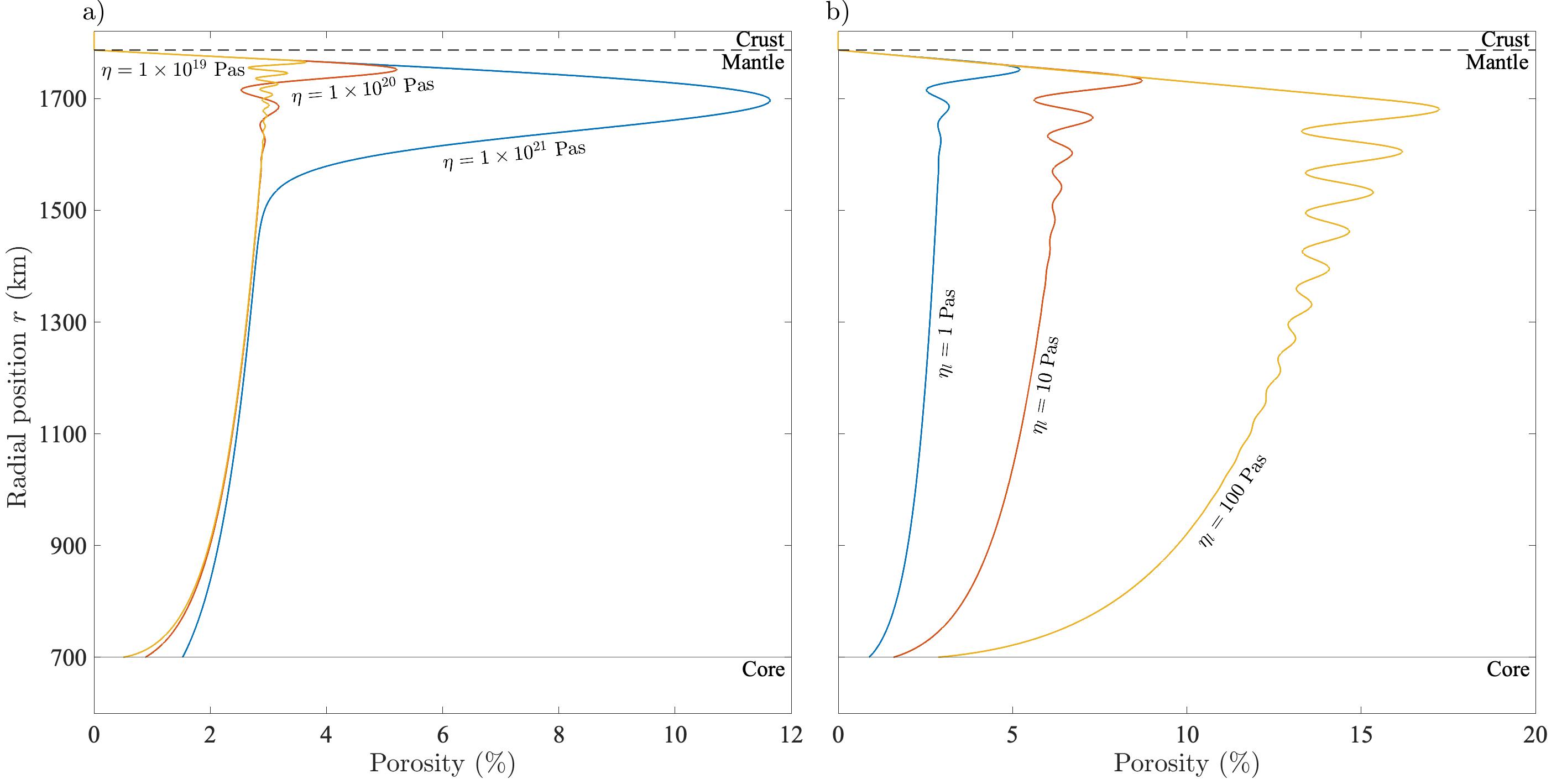}
    \caption{Porosity distribution for varying a) shear viscosity and b) melt viscosity. High shear viscosities cause thick decompacting boundary layers with high peak porosities. High melt viscosities cause high-melt-fractions throughout the mantle as melt can less easily segregate. High shear viscosity and low melt fractions increase the bulk viscosity, damping porosity oscillations.}
    \label{eta_vary}
\end{figure}

We now examine the effects of varying material and rheological parameters in the model. Figure \ref{Pc_vary} shows solutions of porosity and pressure for a range of critical compaction pressures $P_{c}$, and figure \ref{eta_vary} shows solutions of porosity for a range of matrix shear viscosities $\eta$ and liquid visosities $\eta_{l}$. We consider the critical overpressure $P_{c}$ to be a material parameter as it parameterises the strength of the downwelling crust.

Pressure differences between the solid and liquid in the partially molten rock causes the solid matrix to deform. When the solid pressure is higher, the matrix compacts, expelling liquid, and when the liquid pressure is higher, the solid decompacts and melt accumulates. Away from boundary layers, the melt fraction is almost entirely controlled by the buoyant segregation of magma. Figures \ref{Pc_vary} and \ref{eta_vary}a show that $P_{c}$ and $\eta$ do not affect the melt fraction outside of boundary layers, while figure \ref{eta_vary}b shows that increasing the liquid viscosity $\eta_{l}$ increases melt fractions throughout the mantle. This is because the buoyancy-driven melt fraction is controlled by the permeability and liquid viscosity in Darcy's law (\ref{eq:Darcy}).

As solid crust downwells, warms, and begins to melt, the high pressure of rising magma forces the solid matrix to decompact to accommodate infiltration of buoyant magma. This decompaction occurs over a region known as a decompacting boundary layer, in which the compaction pressure gradient term in Darcy's law (\ref{eq:Darcy}) becomes important. Figure \ref{Pc_vary} shows that if a large compaction pressure (liquid overpressure) is required for magma to move out of the mantle pore-space into the plumbing system, large porosities build up beneath the crust. This is because the large liquid overpressure drives rapid decompaction of the downwelling solid material. High shear viscosities also lead to the development of a larger boundary layer with higher peak porosity (fig.~\ref{eta_vary}a). An increased shear viscosity (and so through $\zeta=\eta/\phi$, an increased bulk viscosity) causes greater resistance of the downwelling matrix to compaction, which means a larger length scale over which the compaction pressure gradient counteracts buoyancy, and in which the porosity must therefore increase to enable upward melt motion. The thickness of the boundary layer is on the order of the compaction length, which is an emergent length-scale governing the interaction of liquid and solid in a two-phase medium (see appendix B) \cite{mckenzie_generation_1984}.

The interaction between compaction pressure driving melt flow and compaction sets up an oscillation of the porosity and pressure \cite{spiegelman_flow_1993} that decays to match the buoyancy-dominated mantle region below. Figure \ref{eta_vary} shows that high shear viscosities and low magma viscosities (both of which represent high bulk viscosities) damp porosity oscillations effectively so that porosity rapidly relaxes to the buoyancy-driven profile.

In figures \ref{Pc_vary} and \ref{eta_vary} the location of the crust--mantle boundary and the flux of melt out of the mantle are not affected by the parameter changes (where porosity increases, melt velocity decreases and so the flux is unchanged). As such, the material parameters control the form of the decompacting boundary layer but do not affect the crust or volcanic plumbing system. In particular, and perhaps surprisingly, the value of $P_{c}$, which encodes information about the strength of the crust in this model, does not significantly affect the crustal thickness.


\section{Discussion}
Our results demonstrate the importance of magmatic intrusions (emplacement) in controlling Io's crustal thickness and temperature profile. A significant proportion of magmas generated in the mantle must contribute to heating the cold downwelling crust --- otherwise it would be extremely thick. Further, the intrusions must be concentrated in the lower crust if the upper crust is to retain significant elastic strength, which it requires to support Io's high mountains \cite{mckinnon_chaos_2001}. The results also suggest that a high-melt-fraction layer can develop beneath Io's crust due to decompaction, without requiring any radial partitioning of tidal heat. We now discuss these results further and consider their implications.

\subsection{Magmatic intrusions}
\label{magmagtic_intrusion}
The foremost result of this work is captured in figure \ref{h-te-plots}, which indicates that magmatic intrusions are the primary control on Io's crustal thickness. Models of heat piping that neglect heating from intrusions will result in steady-state crustal thicknesses of over $600$~km. Models with a downwelling crust that fix the crustal thickness cannot be assumed to be in thermal steady state and may violate energy conservation. The primary control on the temperature profile in Io's crust is the distribution of magmatic emplacement. Unless intrusions are confined to the lower crust, the upper crust becomes hot and weak, so would be unable to support significant topography. With a sufficiently large elastic-limit temperature, colder regions (in the upper crust) have no emplacement and so remain strong. \citeA{kirchoff_effects_2020} note that the release of confining stress by crustal faults leads to extension in Io's upper crust, which manifest as rifts, pull-apart basins, and simple graben structures. This transition to an extensional regime may further explain the low level of emplacement that must be taking place in the upper crust, with magmas instead rising all the way to the surface. A more detailed description of heat piping might attempt to account for the mechanics and energetics of emplacement in the context of the crustal stress profile.

Unless the crust is unrealistically thin, our model shows that conductive heat loss at the surface is negligible (fig. \ref{h-te-plots}c), consistent with the conclusions of \citeA{oreilly_magma_1981}. As shown in more detail in appendix C, the heat loss due to eruption is $q_{s}(\rho L + \rho c (T_{m}-T_{s}))$ per unit surface area, where $q_{s}$ is the globally averaged eruption rate, $\rho L$ is the latent heat of erupted magma, and $\rho c (T_{m}-T_{s})$ is the sensible heat lost as erupted magma cools to the surface temperature $T_{s}$. At steady state, with negligible conduction through the elastic crust, this surface heat flux must balance the total dissipation rate $\Psi$, implying a volcanic resurfacing rate
\begin{linenomath*}
\begin{equation}
4 \pi R^{2}q_{s} = \frac{\Psi}{\rho L+\rho c(T_{m}-T_{s})}.
\label{eq:erupt_rate}
\end{equation}
\end{linenomath*}
For Io, equation (\ref{eq:erupt_rate}) predicts a resurfacing rate of $1.25~$cm/yr \cite{breuer_10.08_2015}; this is the maximum rate seen in figure \ref{h-te-plots}b. Equation (\ref{eq:erupt_rate}) also provides a means of estimating eruption rates for other tidally heated lava-worlds, utilising their tidal heating rate, size, and surface temperature, all of which are obtainable from observations \cite{bolmont_tidal_2013}. In particular we note the potential application of equation (\ref{eq:erupt_rate}) --- and the model in general --- to the TRAPPIST-1 planets, which are undergoing comparable levels of tidal heating with moderate surface temperatures (150--400~K) \cite{barr_interior_2018}.

The predicted eruption rate can be compared to an estimate of the total melt production rate. Assuming that all tidal heating directly causes melting, the total melt production rate is $\Psi/\rho L$. The predicted eruption rate is therefore less than the melt production rate by a factor of $c(T_{m}-T_{s})/(L+c(T_{m}-T_{s}))$. For Io, this indicates that around $80\%$ of magma must be emplaced into the crust. This analysis demonstrates that the eruption rate and the total amount of emplaced material are controlled by the tidal heating rate and the relative temperatures of the magma and the surface. This explains why these quantities remain almost unchanged in figures \ref{hhat_vary} and \ref{Te_vary}, despite significant differences in the thickness of the crust and the precise location at which emplacement occurs. Equation (\ref{eq:erupt_rate}) also shows that if the total heating rate $\Psi$ were increased, it results in a proportional increase in both the eruption rate and the intrusive emplacement rate.  We find that, for fixed values of the emplacement rate parameters $h$ and $T_{e}$, such an increase results in an increase to the predicted steady-state crustal thickness.

\subsection{Magma-rich layer beneath the crust}
Figures \ref{Pc_vary} and \ref{eta_vary} demonstrate that if a high-melt-fraction layer exists within Io, it does not necessarily imply the presence of a magma ocean with melt fractions above disaggregation \cite{khurana_evidence_2011,tyler_tidal_2015}, nor was it necessarily formed by concentrated tidal heating \cite{moore_thermal_2001,bierson_test_2016}. Instead, our model suggests that a high-melt-fraction layer could form if a large liquid overpressure is required in order to inject dikes into the crust, or if the compaction length is large.

\citeA{mckinnon_chaos_2001} and \citeA{kirchoff_formation_2009} note that high compressive stresses must arise in Io's crust due to the downwelling of a spherical shell. If these compressive stresses are present at the crust--mantle boundary, they will have to be overcome by magma pressure to form dikes; this would indicate a high value for $P_{c}$, promoting the formation of a high-melt-fraction region beneath the crust. However, we have demonstrated that Io's lower crust must be hot due to extensive magmatic intrusion, and so it is likely that the stresses associated with downwelling crust would be accommodated by faulting in the upper crust and viscous creep in the lower crust. The relevant value for $P_{c}$ in the model is thus unclear; more investigation is needed into the strength and deformation mechanisms of Io's lower crust.

Nevertheless, we have demonstrated that a high-melt-fraction layer can in principle arise from a state of uniform tidal heating. Further work is needed to ascertain whether such a layer is likely to exist. In particular, we must determine whether the required high shear viscosities are compatible with the observed tidal heating \cite{bierson_test_2016,renaud_increased_2018}. If such a layer does not exist, our model suggests that melt fractions are relatively uniform within Io (fig. \ref{eta_vary}), providing little drive for radial partitioning of tidal heating. We note however that the high observed equatorial heat fluxes have been used to infer increased asthenospheric heating \cite{cantrall_variability_2018}. If melt fractions are indeed uniform, it is unclear how these increased equatorial heat fluxes would occur.

We emphasise that the presence and form of the decompacting boundary layer are independent of the model predictions for crustal thickness. In the context of our model, a decompacting boundary layer is a feature beneath the crust that does not affect the mass or heat fluxes out of the mantle, nor how energy is deposited in the crust; it is these factors that control Io's crustal thickness.

\subsection{Model limitations}
A primary simplification made in formulating our model is the assumption of spherical symmetry. Other models suggest that tidal heating does not just vary with radius but also with latitude and longitude. \citeA{steinke_tidally_2020} show that if dissipation is significantly concentrated in a layer beneath the crust, lateral variations in mantle temperature can exceed $100~$K, perhaps impacting the usefulness of a one-dimensional approach. More distributed heating leads to much lower lateral mantle temperature variations ($\sim$ $1~$K, \citeA{steinke_tidally_2020}), but may still cause significant lateral variations in melting rate. The presence or absence of a high-melt-fraction layer is thus key to understanding Io's three-dimensional tidal heating distribution.

Convection of partially molten mantle is also a potentially important three-dimensional effect. If buoyant segregation is inefficient at transporting heat, convection of the two phase medium could occur and imprint deviations from the one-dimensional model presented here. However, we stress that convection cannot dominate Io's mantle heat transport due to its low efficiency at melt fractions relevant for Io \cite{breuer_10.08_2015}. Also neglected here is that compaction effects can cause the lateral migration and focusing of melt \cite{sparks_melt_1991,turner_magmatic_2017}, which could lead to channelisation of melt and may exert a control on Io's volcano distribution. Our model represents a long-term average resurfacing rate that is spatially uniform. Volcanic eruptions are discrete events in both space and time, so at any given instant the eruption and resurfacing rate will not be spatially uniform. Our model does not resolve the details of the eruption process, which are likely to include shorter time-scale variability in both space and time around the average rate. As such some caution is required in comparing detailed observations of the current surface heat flux and eruption rate with that predicted by the model.

Another key simplification in this model is the assumption that Io is composed of a single chemical component. Continued melting in the interior is likely to have caused significant chemical stratification \cite{keszthelyi_magmatic_1997}. Melting of a polymineralic rock occurs over a range of temperatures, with more fusible minerals melting first. The upward migration and eruption of fusible melts plausibly depletes the deep mantle of fusible material and enriches the near-surface. Interesting is the lack of observed olivine in surface erupta \cite{keszthelyi_post-galileo_2004}, indicating either that deep refractory melts do not form, or that they predominantly freeze in the interior. Erupted fusible material would also melt at shallow depths upon burial, affecting the crustal thickness. The dynamics of this segregation is omitted here and may exert important controls on Io's volcanism and mantle melt distribution.

\section{Conclusions}
Io is a body that is complicated in its detail but observations suggest it has a simple structure at leading order. We have demonstrated that a coupled model of magmatic segregation, compaction, and heat-piping can explain this leading-order structure and the associated observations: globally averaged elastic thickness, eruption rate, and surface heat flux, as well as a possible high-melt-fraction layer beneath the crust. We have shown that magmatic intrusions into Io's crust are a fundamental control on its crustal thickness. Without the heating associated with the formation of magmatic intrusions, the crust would grow to be $>600$~km thick. However, these intrusions must be confined to the lower crust if the upper crust is to retain sufficient strength to support Io's high mountains. We have also shown that an inferred high-melt-fraction region can be understood as a decompacting boundary layer if a process such as lateral compression makes it difficult for magma to migrate from the mantle into the crust. An extension of our model to include more elaborate chemical thermodynamics and three-dimensional flows will give insights into deviations from the spherically symmetric model developed here.

\appendix
\section{Scaled Model and Non-dimensional Parameters}
Here we non-dimensionalise the governing equations. Dimensional parameter definitions are given in table \ref{table:parameters1} in the main text, and the scales and definitions of the non-dimensional parameters are given in table \ref{table:parameters2}. We write for example $u = u_{0}\hat{u}$, where $u_{0}$ is the velocity scale and $\hat{u}$ is the dimensionless velocity, insert similar expressions for all the variables into the equations, and finally drop the hats on the dimensionless quantities to arrive at a dimensionless model. For the temperature we write $T=T_{s}+T_{0}\hat{T}$ ~with $T_{0}=T_{m}-T_{s}$ so that the non-dimensional temperature varies between $0$ and $1$. We also assume spherical symmetry, and write all quantities as a function of $r$, noting that $\bm{\nabla}\cdot \bm{u} = r^{-2} ~\partial(r^{2}u)/\partial r $ where $u$ is the radial component of the solid velocity.

\begin{table}
\caption{Reference scales and non-dimensional parameters}
\centering
\begin{tabular}{l l l l l}
\hline
Quantity & Symbol & Definition & Preferred Value & Units \\
\hline
Tidal heating scale & $\psi_{0}$ & & $4.2\times 10^{-6}$ & W/m$^{3}$ \\
Liquid velocity scale & $q_{0}$ & $q_{0} = \psi_{0}R/\rho L $ & $6.4\times 10^{-9}$ & m/s \\
Solid velocity scale & $u_{0}$ & $u_{0} = q_{0} $ & $6.4\times 10^{-9}$ & m/s \\
Porosity scale & $\phi_{0}$ & $q_{0} = K_{0}\phi_{0}^{n}\Delta \rho g/\eta_{l}$ & $0.044$ & \\
Temperature scale & $T_{0}$ & $T_{0}=T_{m}-T_{s}$ & $1550$ & K \\
Bulk viscosity scale & $\zeta_{0}$ & $\zeta_{0}=\eta/\phi_{0}$ & $2.3\times 10^{21}$ & Pas \\
Pressure scale & $P_{0}$ & $P_{0}=\zeta_{0}q_{0}/R$ & $8.0\times 10^{6}$ & Pa \\
 & & & & \\
P\'eclet Number & Pe & Pe~$=q_{0}R/\kappa$ & $1160$ & \\
Stefan Number & St & St~$=L/cT_{0}$ & $0.25$ & \\
Emplacement constant & $\hat{h}$ & $\hat{h}=h\rho cT_{0}/\psi_{0}$ &$200$ & \\
Extraction constant & $\hat{\nu}$ & $\hat{\nu}=\nu \zeta_{0}$ & $1000$ & \\
Scaled elastic limit temperature & $\hat{T}_{e}$ & $\hat{T}_{e} =\frac{T_{e}-T_{s}}{T_{m}-T_{s}}$ & $0.6$ & \\
Compaction parameter & $\delta$ & $\delta = \zeta_{0}K_{0}\phi_{0}^{n}/\eta_{l} R^{2} \quad$ & $5.8\times 10^{-3}$ & \\
\hline
\end{tabular}\\
{\raggedright The tidal heating scale $\psi_{0}$ is imposed, which gives the velocity scale $q_{0}$ which in turn gives the porosity scale $\phi_{0}$. \par}
\label{table:parameters2}
\end{table}

The non-dimensional equations for conservation of solid and liquid mass are
\begin{linenomath*}
\begin{equation}
\frac{\partial}{\partial t}(1-\phi_{0}\phi) + \frac{1}{r^{2}}\frac{\partial}{\partial r}\left(r^{2}(1-\phi_{0}\phi)u\right) = - \Gamma + M,
\label{eq:Smass_nondim}
\end{equation}
\end{linenomath*}
\begin{linenomath*}
\begin{equation}
\phi_{0}\frac{\partial\phi}{\partial t}+ \frac{1}{r^{2}}\frac{\partial}{\partial r}\left(r^{2}(\phi_{0}\phi u + q)\right) = \Gamma - E,
\label{eq:Lmass_nondim}
\end{equation}
\end{linenomath*}
where the emplacement rate $M=\hat{h}(1-T)\mathcal{I}_{M}$. $\mathcal{I}_{M}$ is an indicator function that equals $1$ for $T>T_{e}$ and $q_{p}>0$, and equals zero otherwise. This ensures that emplacement only occurs above the elastic limit temperature, and provided there is melt present in the plumbing system to be emplaced. $E=\hat{\nu}(P-P_{c})\mathcal{I}_{E}$ is the extraction rate, where the indicator function $\mathcal{I}_{E}$ equals $1$ for $P>P_{c}$, and equals zero otherwise, ensuring that extraction only occurs in regions above the critical overpressure.

Total conservation of mass for the crust--mantle and plumbing system are
\begin{linenomath*}
\begin{equation}
\frac{1}{r^{2}}\frac{\partial}{\partial r}(r^{2}(u+q)) = M - E,
\label{eq:M_cont_scaled}
\end{equation}
\end{linenomath*}
\begin{linenomath*}
\begin{equation}
\frac{1}{r^{2}}\frac{\partial(r^{2}q_{p})}{\partial r} = E - M.
\label{eq:M_plum_scaled}
\end{equation}
\end{linenomath*}

Darcy's law and the compaction relation become
\begin{linenomath*}
\begin{equation}
q = \phi^{n}\left(1-\phi_{0}\phi-\delta \frac{\partial P}{\partial r}\right), \quad \quad \phi P = \frac{1}{r^{2}}\frac{\partial(r^{2}u)}{\partial r},
\label{eq:darcy_nondim}
\end{equation}
\end{linenomath*}
where $\delta$ is a dimensionless parameter defined in table \ref{table:parameters2}. This measures the typical size of the compaction pressure gradients relative to the buoyancy; it is expected to be relatively small. It can be related to the compaction length \cite{mckenzie_generation_1984} $l = \sqrt{\zeta_{0}K_{0}\phi_{0}^{n}/\eta_{l}}$, by $\delta = l^{2}/R^{2}$ (so the square root of $\delta$ is the ratio of the compaction length to the radius of the planet).

Conservation of energy becomes
\begin{linenomath*}
\begin{equation}
\frac{\partial H}{\partial t} + \frac{1}{r^{2}}\frac{\partial}{\partial r}(r^{2}(u+q)T) + \frac{\textrm{St}}{r^{2}}\frac{\partial}{\partial r}(r^{2}(\phi_{0}\phi u + q)) = \frac{1}{\textrm{Pe}~ r^{2}}\frac{\partial}{\partial r}\left( r^{2}\frac{\partial T}{\partial r}\right) + \textrm{St} \psi + M(1+\textrm{St}) - E(T+\textrm{St}).
\label{eq:E_scaled}
\end{equation}
\end{linenomath*}

\section{Asymptotic Approximation}
To facilitate a rapid exploration of parameter space, we construct an approximation to the steady states of the model. This approximation makes use of the fact that the porosity scale $\phi_{0}$ and the compaction parameter $\delta$ are both much less than unity. Neglecting them in the equations provides a good approximation over most of the crust and the mantle, apart from in the boundary layer just below the crust--mantle boundary (which is discussed below). Note that $\textrm{Pe}^{-1}$ is also expected to be a small parameter, but we retain conduction in the equations because it is important in controlling the temperature distribution within the crust.

\subsection{Mantle and Crust}
In this approximation we consider the crust and mantle separately and solve for the position of the boundary $r=r_{c}$ between them. We assume that all extraction from the mantle occurs within the decompacting boundary layer, just below the base of the crust (as is verified by the full numerical solutions). The extraction term $E$ is therefore non-zero only in a narrow region of thickness $O(\delta)$ and hence is neglected from the continuum equations. At the base of the crust $r=r_{c}$, the entire flux $q(r_{c})$ (hereafter referred to as $q_{c}$) is transferred into the plumbing system. As $q_{p}=0$ in the mantle, emplacement $M$ only appears in the crustal equations. Taking this into account, the combination of mass continuity equations (\ref{eq:Smass_nondim}) and (\ref{eq:Lmass_nondim}) indicate that $u=-q$ throughout the mantle. Since we also have $T=1$ in the mantle, conservation of energy (eqn. \ref{eq:E_scaled}) becomes simply
\begin{linenomath*}
\begin{equation}
    \frac{1}{r^{2}}\frac{\partial(r^{2}q)}{\partial r} = \psi,
    \label{eq:AA_E}
\end{equation}
\end{linenomath*}
so melting (or equivalently, the transport of latent heat by melt) balances tidal heating. Equation (\ref{eq:AA_E}) can be integrated directly to give
\begin{linenomath*}
\begin{equation}
    q(r) = -u(r) = \frac{\psi}{3}\left(r-\frac{r_{b}^{3}}{r^{2}}\right).
    \label{eq:q_outer}
\end{equation}
\end{linenomath*}
Darcy's law (eqn. \ref{eq:darcy_nondim}a) becomes $q=\phi^{n}$ so $\phi$ is deduced directly from $q$, and the compaction equation (\ref{eq:darcy_nondim}b) further indicates that $P=-\psi/\phi$. In particular, these expressions give the values for the flux $q_{c}$, porosity, and compaction pressure at the top of the mantle $r=r_{c}$, which are used below to feed into the decompacting boundary layer,
\begin{linenomath*}
\begin{equation}
    q_{c} = \frac{\psi}{3}\left(r_{c}-\frac{r_{b}^{3}}{r_{c}^{2}}\right), \quad\quad \phi(r_{c}) = q_{c}^{1/n}, \quad\quad P(r_{c}) = \frac{\psi}{q_{c}^{1/n}}.
    \label{eq:into_DBL}
\end{equation}
\end{linenomath*}
The flux $q_{c}$ transfers to the plumbing system at the top of the mantle for its continued transport through the crust, to which we now turn.

In the crust, $q=0$ and the combination of mass equations (\ref{eq:M_cont_scaled}) and (\ref{eq:M_plum_scaled}) now requires $u=-q_{p}$, the plumbing-system flux. Conservation of mass in the plumbing system (eqn. \ref{eq:M_plum_scaled}) becomes
\begin{linenomath*}
\begin{equation}
    \frac{1}{r^{2}}\frac{\partial(r^{2}q_{p})}{\partial r}= -M.
    \label{eq:AA_crust_qp}
\end{equation}
\end{linenomath*}
Conservation of energy in the crust (eqn. \ref{eq:E_scaled}) becomes
\begin{linenomath*}
\begin{equation}
    \frac{1}{r{^2}}\frac{\partial}{\partial r}(r^{2}uT) = \frac{1}{\textrm{Pe}~ r^{2}}\frac{\partial}{\partial r}\left( r^{2}\frac{\partial T}{\partial r} \right) + \textrm{St} \psi + M(\textrm{St} +1),
    \label{eq:AA_crust_E}
\end{equation}
\end{linenomath*}
so advection of the solid balances conduction, tidal heating, and heating from intrusions (emplacement). These two equations ((\ref{eq:AA_crust_qp}) and (\ref{eq:AA_crust_E})) are solved together to determine the temperature profile $T$, the plumbing system flux $q_{p}$ (and hence the solid velocity $u$), and the position of the crust-mantle boundary $r_{c}$. The required boundary conditions are
\begin{linenomath*}
\begin{equation}
\begin{split}
q_{p}=q_{c}, \quad T = 1, \quad \frac{\partial T}{\partial r}=0, \quad & \textrm{at}~ r = r_{c}, \\
T = 0, \quad & \textrm{at}~ r = 1.
\end{split}
\end{equation}
\end{linenomath*}
Although this solution to the crustal system involves a numerical integration, it is considerably more straightforward and faster than the solution to the full model.

From this approach we find a good approximation to the thickness of the crust, the temperature profile within the crust, the plumbing flux $q_{p}$ and emplacement rate (and hence the eruptive flux at the surface), as well as the porous melt flux and porosity in the majority of the mantle. A detail missing from the full solutions that is not yet captured by this asymptotic approximation is the high-porosity region --- the decompacting boundary layer --- just below the base of the crust. In the context of our model, apart from transferring melt from the porous mantle to the plumbing system, the details of this layer are unimportant in determining the large scale structure of the solutions (thickness and temperature distribution of the crust). However, in order to understand the dynamics further, we now analyse this region.

\subsection{Decompacting Boundary Layer}
The behaviour in the boundary layer is obtained by rescaling the equations locally to find a local approximation of the solution close to the crust--mantle boundary. A composite approximation valid over the whole domain can then be found by combining the two approximations (the `outer' mantle and crust solution given above, and the `inner' boundary layer solution).

As rising magma approaches the crust--mantle boundary at $r=r_{c}$, it is no longer reasonable to neglect the compaction pressure gradient term in equation (\ref{eq:darcy_nondim}). Approaching the crust, rising magma is impeded by the low permeability of downwelling solid, which causes magma to accumulate. The accumulation of magma in this layer generates pressure that decompacts the low-porosity-solid that is downwelling from the crust (cf.~\citeA{hewitt_partial_2008}). To understand what happens in this boundary layer, we must reintroduce the term proportional to $\delta$ in equation (\ref{eq:darcy_nondim}), and rescale lengths to consider the dynamics close to the crust--mantle boundary.

First we note that including extraction, the conservation of mass equations for the solid, liquid, and plumbing system in the mantle are
\begin{linenomath*}
\begin{equation}
    \frac{1}{r^{2}}\frac{\partial}{\partial r}\left( r^{2}(1-\phi_{0}\phi)u\right) = -\psi, \quad \frac{1}{r^{2}}\frac{\partial}{\partial r}\left( r^{2}(\phi_{0}\phi u +q)\right) = \psi - E, \quad \frac{1}{r^{2}}\frac{\partial(r^{2}q_{p})}{\partial r}=E.
    \label{eq:DBL_mass}
\end{equation}
\end{linenomath*}
We also note that the compaction pressure relation (eqn. \ref{eq:darcy_nondim}b) can be combined with conservation of solid mass in the mantle (eqn. \ref{eq:DBL_mass}a), giving
\begin{linenomath*}
\begin{equation}
\phi_{0}u\frac{\partial\phi}{\partial r} - (1-\phi_{0}\phi)\phi P = \psi.
\label{eq:combined_nondim_eq}
\end{equation}
\end{linenomath*}
We then write $r=r_{c}-\delta Z$, where $Z$ is a boundary layer coordinate describing the rescaled distance beneath the crust--mantle boundary. Rewriting equations (\ref{eq:DBL_mass}), (\ref{eq:darcy_nondim}a), and (\ref{eq:combined_nondim_eq}) in terms of this coordinate, we find
\begin{linenomath*}
\begin{eqnarray}
\frac{\partial u}{\partial Z}=O(\phi_{0},\delta), \quad \frac{\partial q}{\partial Z}=-E +O(\phi_{0},\delta), \quad \frac{\partial q_{p}}{\partial Z} = E + O(\phi_{0},\delta), \nonumber\\
q = \phi^{n}\left(1+\frac{\partial P}{\partial Z}\right)+O(\phi_{0}), \quad -\mu u\frac{\partial \phi}{\partial Z}=\psi + \phi P + O(\phi_{0},\delta),
\label{eq:DBL_order_eqs}
\end{eqnarray}
\end{linenomath*}
where $O(\phi_{0},\delta)$ represents terms of order $\phi_{0}$ or $\delta$, which may be neglected (note that having rescaled into the boundary layer, the neglected terms are not the same as those neglected earlier, reflecting the different dominant physics in the boundary layer). We write $\phi_{0}/\delta = \mu$, and treat this as an $O(1)$ parameter.

The first mass conservation equation (\ref{eq:DBL_order_eqs}a) indicates that $u$ is approximately constant throughout this layer, and its value is given by matching the value in the mantle below, $u=-q_{c}$, where $q_{c}$ is the liquid flux defined in equation (\ref{eq:into_DBL}). Moreover, adding together the second two mass conservation equations (\ref{eq:DBL_order_eqs}b--c) shows that $q+q_{p}$ is constant, and must be equal to $q_{c}$ to match with the mantle. The final two equations (\ref{eq:DBL_order_eqs}d--e) (which represent Darcy's law and the compaction relation) can therefore be written as
\begin{linenomath*}
\begin{equation}
    q_{c}\mu \frac{\partial \phi}{\partial Z}=\psi + \phi P, \quad \frac{\partial P}{\partial Z}=\frac{q}{\phi^{n}}-1.
    \label{eq:DBL_phase_plane}
\end{equation}
\end{linenomath*}

Extraction occurs over the region $0<Z<Z_{E}$ ($Z_{E}$ is determined shortly). In this region we have $E=\hat{\nu}\delta(P-P_{c})$, and we take the limit of very large extraction rate constant $\nu$ (so $\hat{\nu}\delta$ is large), such that $P\simeq P_{c}$ throughout this region. Equation (\ref{eq:DBL_phase_plane}a) can then be integrated to give
\begin{linenomath*}
\begin{equation}
    \phi(Z) = \frac{\psi}{P_{c}}\left(e^{\frac{P_{c}Z}{q_{c}\mu}} -1\right), \quad \quad 0<Z<Z_{E},
    \label{eq:DBL_phi_analytical}
\end{equation}
\end{linenomath*}
and as $P$ is approximately constant, equation (\ref{eq:DBL_phase_plane}b) gives $q=\phi^{n}$. The porosity and flux $q$ therefore increase with $Z$ through this extraction region and since $q_{p}=q_{c}-q$, the plumbing flux correspondingly decreases with $Z$ until it reaches $0$. This defines the position $Z_{E}$ at which extraction started. Substituting $q_{c}=q=\phi^{n}$ into equation (\ref{eq:DBL_phi_analytical}) gives
\begin{linenomath*}
\begin{equation}
    Z_{E} = \frac{q_{c}\mu}{P_{c}}~\textrm{ln}\left(\frac{P_{c}q_{c}^{1/n}}{\psi} + 1\right).
\end{equation}
\end{linenomath*}

Turning to the region below extraction where $E=0$, mass conservation equations (\ref{eq:DBL_mass}b--c) indicate that $q_{p}=0$ and $q=q_{c}$ are now approximately constant. In this region equations (\ref{eq:DBL_phase_plane}a) and (\ref{eq:DBL_phase_plane}b) comprise a two-dimensional phase-plane problem for $\phi(Z)$ and $P(Z)$. A solution is sought with $P=P_{c}$ and $\phi=q_{c}^{1/n}$ at $Z_{E}$ (for continuity with the extraction region), and that matches the correct far-field behaviour as $Z\rightarrow \infty$. The correct behaviour is that $\phi$ and $P$ tend towards the values $q_{c}^{1/n}$ and $-\psi/q_{c}^{1/n}$, given earlier in equation (\ref{eq:into_DBL}), to match with the rest of the mantle. Since this corresponds to a fixed point of the system that is a stable spiral or node, such a solution can be found. The solution involves decaying oscillations of both $\phi$ and $P$ towards the far-field values, which are evident in figures \ref{hhat_vary} and \ref{Te_vary}, and even more so in figures \ref{Pc_vary} and \ref{eta_vary}.

It it worth pointing out that the $1/\phi$ dependence of bulk viscosity is an important control on the porosity oscillation within this boundary layer. Other forms of bulk viscosity with a weaker singularity have also been suggested; for example with $\zeta \sim -\ln{\phi}$ as $\phi \rightarrow 0$ \cite{rudge_viscosities_2018}. The weaker dependence on porosity leads to greater oscillations, but the general form of the boundary layer is maintained.

To compare the asymptotic approximation with the full numerical solution in figures \ref{hhat_vary} and \ref{Te_vary}, we construct a combination of the `outer' solution for the majority of the mantle (where $q$ is given by equation (\ref{eq:q_outer}), $\phi=q^{1/n}$, and $P=\psi/\phi$), together with the `inner' solutions for the decompacting boundary layer. Denoting the former solution as $\phi_{m}(r)$ and the latter solution $\phi_{bl}(Z)$, this composite solution is defined by
\begin{linenomath*}
\begin{equation}
    \phi(r) = \phi_{m}(r) + \phi_{bl}\left(\frac{r_{c}-c}{\delta}\right) - \phi_{l},
\end{equation}
\end{linenomath*}
with equivalent expressions for $P$ and $q$, where the subtraction of the `overlapping' value $\phi_{l}$ is necessary to avoid double counting. The solutions so obtained are re-dimensionalised to produce the approximate solution that is shown as the dashed lines in figures \ref{hhat_vary} and \ref{Te_vary}, which is in good agreement with the full solutions.

\section{Analysis of Heat Flux and Emplacement}
Useful information can be obtained by integrating the energy equation (\ref{eq:AA_crust_E}) over the crust (from $r=r_{c}$ to $r=1$). Substituting equation (\ref{eq:AA_crust_qp}) into equation (\ref{eq:AA_crust_E}), recalling that $u=-q_{p}$ and integrating, gives
\begin{linenomath*}
\begin{equation}
    \left((\textrm{St} +1)q_{p} - \frac{1}{\textrm{Pe}}\frac{\partial T}{\partial r}\right)\bigg| _{r=1} = \textrm{St} r_{c}^{2}q_{c} + \textrm{St} \int_{r_{c}}^{1}\psi~r^{2}{\rm d}r
    \label{eq:crust_int1}
\end{equation}
\end{linenomath*}
where $q_{c}$ is the melt flux at the top of the mantle, which was defined in equation (\ref{eq:into_DBL}). Equation (\ref{eq:crust_int1}) can thus be written as
\begin{linenomath*}
\begin{equation}
    \left((\textrm{St} +1)q_{p} - \frac{1}{\textrm{Pe}}\frac{\partial T}{\partial r}\right)\bigg| _{r=1} = \textrm{St} \int_{r_{b}}^{1}r^{2}\psi {\rm d}r.
\end{equation}
\end{linenomath*}
where we recall that $r_{b}$ is the radius of the core. The left hand side here represents the heat loss due to eruption and conduction at the surface, and the right hand side is the total tidal input; this expression thus represents a global energy balance.

After re-dimensionalising the variables, this can be written as
\begin{linenomath*}
\begin{equation}
4\pi R^{2}\left(q_{p}\left(\rho L + \rho cT_{m}-\rho cT_{s}\right) - k\frac{\partial T}{\partial r}\right)\bigg| _{r=R} = \Psi,
\end{equation}
\end{linenomath*}
where $k$ is the conductivity.

As shown in figure \ref{h-te-plots}, for the parameter regime applicable to Io, the surface conductive heat flux is negligible. In this case we see that the resurfacing rate $q_{s}$ (that is, the value of $q_{p}$ at the surface) is given by
\begin{linenomath*}
\begin{equation}
q_{s} = \frac{\Psi}{4\pi R^{2} (\rho L + \rho c(T_{m}-T_{s}))}.
\label{lava-world-volcanism}
\end{equation}
\end{linenomath*}
This expression can be contrasted with the liquid flux into the crust $q_{c}$ in equation \ref{eq:into_DBL}. Since $r_{c}$ is typically close to $R$, that expression can be written dimensionally as
\begin{linenomath*}
\begin{equation}
q_{c} \approx \frac{\Psi}{4\pi R^{2}\rho L}.
\label{lava-world-base}
\end{equation}
\end{linenomath*}
As such, $q_{s}$ is approximately a fraction $L/[L+c(T_{m}-T_{s})]$ of $q_{c}$, and the proportion of $q_{c}$ that must be emplaced is
\begin{linenomath*}
\begin{equation}
\frac{c(T_{m}-T_{s})}{L + c(T_{m}-T_{s})}.
\label{eq:emplace_frac}
\end{equation}
\end{linenomath*}
For Io, equation (\ref{lava-world-volcanism}) gives a resurfacing rate of of $1.25~$cm/yr. Compared to the flux into the base of the crust (eqn. \ref{lava-world-base}) of $6.3~$cm/yr, equation (\ref{eq:emplace_frac}) predicts that $80\%$ of magma produced inside Io is emplaced into the crust.

%
%
%
%
%
%
%
%

\acknowledgments
This work was funded by the \textit{Science and Technologies Facilities Council}, the University of Oxford's \textit{Oxford Radcliffe Scholarship}, and \textit{University College, Oxford}. This research received funding from the European Research Council under the European Union’s Horizon 2020 research and innovation programme grant agreement number 772255. We thank D. Stevenson, W. Moore, P. England, J. Wade, J. Spencer, J. Rudge, O. Shorttle, D. May and L. Montesi for their comments and suggestions. Source code and data used in the production of figures can be found in \citeA{spencer_spencer-spaceio_compaction_2020}.


%
%

\bibliography{Oxford}

\begin{thebibliography}{}

\bibitem [\protect \citeauthoryear {%
Balay%
\ \protect \BOthers {.}}{%
Balay%
\ \protect \BOthers {.}}{%
{\protect \APACyear {2019}}%
{\protect \APACexlab {{\protect \BCnt {1}}}}}]{%
petsc-user-ref}
\APACinsertmetastar {%
petsc-user-ref}%
\begin{APACrefauthors}%
Balay, S.%
, Abhyankar, S.%
, Adams, M\BPBI F.%
, Brown, J.%
, Brune, P.%
, Buschelman, K.%
\BDBL {}Zhang, H.%
\end{APACrefauthors}%
\unskip\
\newblock
\APACrefYearMonthDay{2019{\protect \BCnt {1}}}{}{}.
\newblock
\APACrefbtitle {{PETS}c Users Manual} {{PETS}c users manual}\
  \APACbVolEdTR{}{\BTR{}\ \BNUM\ ANL-95/11 - Revision 3.12}.
\newblock
\APACaddressInstitution{}{Argonne National Laboratory}.
\newblock
\begin{APACrefURL} \url{https://www.mcs.anl.gov/petsc} \end{APACrefURL}
\PrintBackRefs{\CurrentBib}

\bibitem [\protect \citeauthoryear {%
Balay%
\ \protect \BOthers {.}}{%
Balay%
\ \protect \BOthers {.}}{%
{\protect \APACyear {2019}}%
{\protect \APACexlab {{\protect \BCnt {2}}}}}]{%
petsc-web-page}
\APACinsertmetastar {%
petsc-web-page}%
\begin{APACrefauthors}%
Balay, S.%
, Abhyankar, S.%
, Adams, M\BPBI F.%
, Brown, J.%
, Brune, P.%
, Buschelman, K.%
\BDBL {}Zhang, H.%
\end{APACrefauthors}%
\unskip\
\newblock
\APACrefYearMonthDay{2019{\protect \BCnt {2}}}{}{}.
\newblock
\APACrefbtitle {{PETS}c {W}eb page.} {{PETS}c {W}eb page.}
\newblock
\APAChowpublished {https://www.mcs.anl.gov/petsc}.
\newblock
\begin{APACrefURL} \url{https://www.mcs.anl.gov/petsc} \end{APACrefURL}
\PrintBackRefs{\CurrentBib}

\bibitem [\protect \citeauthoryear {%
Balay%
, Gropp%
, McInnes%
\BCBL {}\ \BBA {} Smith%
}{%
Balay%
\ \protect \BOthers {.}}{%
{\protect \APACyear {1997}}%
}]{%
petsc-efficient}
\APACinsertmetastar {%
petsc-efficient}%
\begin{APACrefauthors}%
Balay, S.%
, Gropp, W\BPBI D.%
, McInnes, L\BPBI C.%
\BCBL {}\ \BBA {} Smith, B\BPBI F.%
\end{APACrefauthors}%
\unskip\
\newblock
\APACrefYearMonthDay{1997}{}{}.
\newblock
{\BBOQ}\APACrefatitle {Efficient Management of Parallelism in Object Oriented
  Numerical Software Libraries} {Efficient management of parallelism in object
  oriented numerical software libraries}.{\BBCQ}
\newblock
\BIn{} E.~Arge, A\BPBI M.~Bruaset\BCBL {}\ \BBA {} H\BPBI P.~Langtangen\
  (\BEDS), \APACrefbtitle {Modern Software Tools in Scientific Computing}
  {Modern software tools in scientific computing}\ (\BPGS\ 163--202).
\newblock
\APACaddressPublisher{}{Birkh{\"{a}}user Press}.
\PrintBackRefs{\CurrentBib}

\bibitem [\protect \citeauthoryear {%
Barr%
, Dobos%
\BCBL {}\ \BBA {} Kiss%
}{%
Barr%
\ \protect \BOthers {.}}{%
{\protect \APACyear {2018}}%
}]{%
barr_interior_2018}
\APACinsertmetastar {%
barr_interior_2018}%
\begin{APACrefauthors}%
Barr, A\BPBI C.%
, Dobos, V.%
\BCBL {}\ \BBA {} Kiss, L\BPBI L.%
\end{APACrefauthors}%
\unskip\
\newblock
\APACrefYearMonthDay{2018}{{\APACmonth{05}}}{}.
\newblock
{\BBOQ}\APACrefatitle {Interior structures and tidal heating in the
  {TRAPPIST}-1 planets} {Interior structures and tidal heating in the
  {TRAPPIST}-1 planets}.{\BBCQ}
\newblock
\APACjournalVolNumPages{Astronomy \& Astrophysics}{613}{}{A37}.
\newblock
\begin{APACrefURL}
  [{2020-04-28}]\url{https://www.aanda.org/articles/aa/abs/2018/05/aa31992-17/aa31992-17.html}
  \end{APACrefURL}
\newblock
\APACrefnote{Publisher: EDP Sciences}
\newblock
\begin{APACrefDOI} \doi{10.1051/0004-6361/201731992} \end{APACrefDOI}
\PrintBackRefs{\CurrentBib}

\bibitem [\protect \citeauthoryear {%
Bierson%
\ \BBA {} Nimmo%
}{%
Bierson%
\ \BBA {} Nimmo%
}{%
{\protect \APACyear {2016}}%
}]{%
bierson_test_2016}
\APACinsertmetastar {%
bierson_test_2016}%
\begin{APACrefauthors}%
Bierson, C\BPBI J.%
\BCBT {}\ \BBA {} Nimmo, F.%
\end{APACrefauthors}%
\unskip\
\newblock
\APACrefYearMonthDay{2016}{{\APACmonth{11}}}{}.
\newblock
{\BBOQ}\APACrefatitle {A test for {Io}'s magma ocean: {Modeling} tidal
  dissipation with a partially molten mantle} {A test for {Io}'s magma ocean:
  {Modeling} tidal dissipation with a partially molten mantle}.{\BBCQ}
\newblock
\APACjournalVolNumPages{Journal of Geophysical Research:
  Planets}{121}{11}{2211--2224}.
\newblock
\begin{APACrefURL}
  \url{http://onlinelibrary.wiley.com/doi/10.1002/2016JE005005/abstract}
  \end{APACrefURL}
\newblock
\begin{APACrefDOI} \doi{10.1002/2016JE005005} \end{APACrefDOI}
\PrintBackRefs{\CurrentBib}

\bibitem [\protect \citeauthoryear {%
Bl{\"o}cker%
, Saur%
, Roth%
\BCBL {}\ \BBA {} Strobel%
}{%
Bl{\"o}cker%
\ \protect \BOthers {.}}{%
{\protect \APACyear {2018}}%
}]{%
blocker_mhd_2018}
\APACinsertmetastar {%
blocker_mhd_2018}%
\begin{APACrefauthors}%
Bl{\"o}cker, A.%
, Saur, J.%
, Roth, L.%
\BCBL {}\ \BBA {} Strobel, D\BPBI F.%
\end{APACrefauthors}%
\unskip\
\newblock
\APACrefYearMonthDay{2018}{}{}.
\newblock
{\BBOQ}\APACrefatitle {{MHD} {Modeling} of the {Plasma} {Interaction} {With}
  {Io}'s {Asymmetric} {Atmosphere}} {{MHD} {Modeling} of the {Plasma}
  {Interaction} {With} {Io}'s {Asymmetric} {Atmosphere}}.{\BBCQ}
\newblock
\APACjournalVolNumPages{Journal of Geophysical Research: Space
  Physics}{123}{11}{9286--9311}.
\newblock
\begin{APACrefURL}
  [{2019-11-11}]\url{https://agupubs.onlinelibrary.wiley.com/doi/abs/10.1029/2018JA025747}
  \end{APACrefURL}
\newblock
\begin{APACrefDOI} \doi{10.1029/2018JA025747} \end{APACrefDOI}
\PrintBackRefs{\CurrentBib}

\bibitem [\protect \citeauthoryear {%
Bolmont%
\ \protect \BOthers {.}}{%
Bolmont%
\ \protect \BOthers {.}}{%
{\protect \APACyear {2013}}%
}]{%
bolmont_tidal_2013}
\APACinsertmetastar {%
bolmont_tidal_2013}%
\begin{APACrefauthors}%
Bolmont, E.%
, Selsis, F.%
, Raymond, S\BPBI N.%
, Leconte, J.%
, Hersant, F.%
, Maurin, A\BHBI S.%
\BCBL {}\ \BBA {} Pericaud, J.%
\end{APACrefauthors}%
\unskip\
\newblock
\APACrefYearMonthDay{2013}{{\APACmonth{08}}}{}.
\newblock
{\BBOQ}\APACrefatitle {Tidal dissipation and eccentricity pumping:
  {Implications} for the depth of the secondary eclipse of 55 {Cancri} e}
  {Tidal dissipation and eccentricity pumping: {Implications} for the depth of
  the secondary eclipse of 55 {Cancri} e}.{\BBCQ}
\newblock
\APACjournalVolNumPages{Astronomy \& Astrophysics}{556}{}{A17}.
\newblock
\begin{APACrefURL}
  [{2019-08-01}]\url{https://www.aanda.org/articles/aa/abs/2013/08/aa20837-12/aa20837-12.html}
  \end{APACrefURL}
\newblock
\begin{APACrefDOI} \doi{10.1051/0004-6361/201220837} \end{APACrefDOI}
\PrintBackRefs{\CurrentBib}

\bibitem [\protect \citeauthoryear {%
Breuer%
\ \BBA {} Moore%
}{%
Breuer%
\ \BBA {} Moore%
}{%
{\protect \APACyear {2015}}%
}]{%
breuer_10.08_2015}
\APACinsertmetastar {%
breuer_10.08_2015}%
\begin{APACrefauthors}%
Breuer, D.%
\BCBT {}\ \BBA {} Moore, W\BPBI B.%
\end{APACrefauthors}%
\unskip\
\newblock
\APACrefYearMonthDay{2015}{{\APACmonth{01}}}{}.
\newblock
{\BBOQ}\APACrefatitle {10.08 - {Dynamics} and {Thermal} {History} of the
  {Terrestrial} {Planets}, the {Moon}, and {Io}} {10.08 - {Dynamics} and
  {Thermal} {History} of the {Terrestrial} {Planets}, the {Moon}, and
  {Io}}.{\BBCQ}
\newblock
\BIn{} G.~Schubert\ (\BED), \APACrefbtitle {Treatise on {Geophysics} ({Second}
  {Edition})} {Treatise on {Geophysics} ({Second} {Edition})}\ (\BPGS\
  255--305).
\newblock
\APACaddressPublisher{Oxford}{Elsevier}.
\newblock
\begin{APACrefURL}
  [{2019-11-11}]\url{http://www.sciencedirect.com/science/article/pii/B9780444538024001731}
  \end{APACrefURL}
\newblock
\begin{APACrefDOI} \doi{10.1016/B978-0-444-53802-4.00173-1} \end{APACrefDOI}
\PrintBackRefs{\CurrentBib}

\bibitem [\protect \citeauthoryear {%
Cantrall%
\ \protect \BOthers {.}}{%
Cantrall%
\ \protect \BOthers {.}}{%
{\protect \APACyear {2018}}%
}]{%
cantrall_variability_2018}
\APACinsertmetastar {%
cantrall_variability_2018}%
\begin{APACrefauthors}%
Cantrall, C.%
, de Kleer, K.%
, de Pater, I.%
, Williams, D\BPBI A.%
, Davies, A\BPBI G.%
\BCBL {}\ \BBA {} Nelson, D.%
\end{APACrefauthors}%
\unskip\
\newblock
\APACrefYearMonthDay{2018}{{\APACmonth{09}}}{}.
\newblock
{\BBOQ}\APACrefatitle {Variability and geologic associations of volcanic
  activity on {Io} in 2001–2016} {Variability and geologic associations of
  volcanic activity on {Io} in 2001–2016}.{\BBCQ}
\newblock
\APACjournalVolNumPages{Icarus}{312}{}{267--294}.
\newblock
\begin{APACrefURL}
  [{2020-05-01}]\url{http://www.sciencedirect.com/science/article/pii/S0019103517305018}
  \end{APACrefURL}
\newblock
\begin{APACrefDOI} \doi{10.1016/j.icarus.2018.04.007} \end{APACrefDOI}
\PrintBackRefs{\CurrentBib}

\bibitem [\protect \citeauthoryear {%
Davies%
, Veeder%
, Matson%
\BCBL {}\ \BBA {} Johnson%
}{%
Davies%
\ \protect \BOthers {.}}{%
{\protect \APACyear {2015}}%
}]{%
davies_map_2015}
\APACinsertmetastar {%
davies_map_2015}%
\begin{APACrefauthors}%
Davies, A\BPBI G.%
, Veeder, G\BPBI J.%
, Matson, D\BPBI L.%
\BCBL {}\ \BBA {} Johnson, T\BPBI V.%
\end{APACrefauthors}%
\unskip\
\newblock
\APACrefYearMonthDay{2015}{{\APACmonth{12}}}{}.
\newblock
{\BBOQ}\APACrefatitle {Map of {Io}'s volcanic heat flow} {Map of {Io}'s
  volcanic heat flow}.{\BBCQ}
\newblock
\APACjournalVolNumPages{Icarus}{262}{}{67--78}.
\newblock
\begin{APACrefURL}
  \url{http://www.sciencedirect.com/science/article/pii/S0019103515003474}
  \end{APACrefURL}
\newblock
\begin{APACrefDOI} \doi{10.1016/j.icarus.2015.08.003} \end{APACrefDOI}
\PrintBackRefs{\CurrentBib}

\bibitem [\protect \citeauthoryear {%
de Kleer%
\ \BBA {} de Pater%
}{%
de Kleer%
\ \BBA {} de Pater%
}{%
{\protect \APACyear {2016}}%
}]{%
de_kleer_time_2016}
\APACinsertmetastar {%
de_kleer_time_2016}%
\begin{APACrefauthors}%
de Kleer, K.%
\BCBT {}\ \BBA {} de Pater, I.%
\end{APACrefauthors}%
\unskip\
\newblock
\APACrefYearMonthDay{2016}{{\APACmonth{12}}}{}.
\newblock
{\BBOQ}\APACrefatitle {Time variability of {Io}'s volcanic activity from
  near-{IR} adaptive optics observations on 100 nights in 2013–2015} {Time
  variability of {Io}'s volcanic activity from near-{IR} adaptive optics
  observations on 100 nights in 2013–2015}.{\BBCQ}
\newblock
\APACjournalVolNumPages{Icarus}{280}{}{378--404}.
\newblock
\begin{APACrefURL}
  \url{http://linkinghub.elsevier.com/retrieve/pii/S0019103516303104}
  \end{APACrefURL}
\newblock
\begin{APACrefDOI} \doi{10.1016/j.icarus.2016.06.019} \end{APACrefDOI}
\PrintBackRefs{\CurrentBib}

\bibitem [\protect \citeauthoryear {%
Hamilton%
\ \protect \BOthers {.}}{%
Hamilton%
\ \protect \BOthers {.}}{%
{\protect \APACyear {2013}}%
}]{%
hamilton_spatial_2013}
\APACinsertmetastar {%
hamilton_spatial_2013}%
\begin{APACrefauthors}%
Hamilton, C\BPBI W.%
, Beggan, C\BPBI D.%
, Still, S.%
, Beuthe, M.%
, Lopes, R\BPBI M\BPBI C.%
, Williams, D\BPBI A.%
\BDBL {}Wright, W.%
\end{APACrefauthors}%
\unskip\
\newblock
\APACrefYearMonthDay{2013}{{\APACmonth{01}}}{}.
\newblock
{\BBOQ}\APACrefatitle {Spatial distribution of volcanoes on {Io}:
  {Implications} for tidal heating and magma ascent} {Spatial distribution of
  volcanoes on {Io}: {Implications} for tidal heating and magma ascent}.{\BBCQ}
\newblock
\APACjournalVolNumPages{Earth and Planetary Science Letters}{361}{}{272--286}.
\newblock
\begin{APACrefURL}
  \url{http://www.sciencedirect.com/science/article/pii/S0012821X12006012}
  \end{APACrefURL}
\newblock
\begin{APACrefDOI} \doi{10.1016/j.epsl.2012.10.032} \end{APACrefDOI}
\PrintBackRefs{\CurrentBib}

\bibitem [\protect \citeauthoryear {%
Hewitt%
\ \BBA {} Fowler%
}{%
Hewitt%
\ \BBA {} Fowler%
}{%
{\protect \APACyear {2008}}%
}]{%
hewitt_partial_2008}
\APACinsertmetastar {%
hewitt_partial_2008}%
\begin{APACrefauthors}%
Hewitt, I.%
\BCBT {}\ \BBA {} Fowler, A.%
\end{APACrefauthors}%
\unskip\
\newblock
\APACrefYearMonthDay{2008}{{\APACmonth{09}}}{}.
\newblock
{\BBOQ}\APACrefatitle {Partial melting in an upwelling mantle column} {Partial
  melting in an upwelling mantle column}.{\BBCQ}
\newblock
\APACjournalVolNumPages{Proceedings of the Royal Society A: Mathematical,
  Physical and Engineering Sciences}{464}{2097}{2467--2491}.
\newblock
\begin{APACrefURL}
  [{2019-07-17}]\url{https://royalsocietypublishing.org/doi/full/10.1098/rspa.2008.0045}
  \end{APACrefURL}
\newblock
\begin{APACrefDOI} \doi{10.1098/rspa.2008.0045} \end{APACrefDOI}
\PrintBackRefs{\CurrentBib}

\bibitem [\protect \citeauthoryear {%
Katz%
}{%
Katz%
}{%
{\protect \APACyear {2008}}%
}]{%
katz_magma_2008}
\APACinsertmetastar {%
katz_magma_2008}%
\begin{APACrefauthors}%
Katz, R\BPBI F.%
\end{APACrefauthors}%
\unskip\
\newblock
\APACrefYearMonthDay{2008}{{\APACmonth{12}}}{}.
\newblock
{\BBOQ}\APACrefatitle {Magma {Dynamics} with the {Enthalpy} {Method}:
  {Benchmark} {Solutions} and {Magmatic} {Focusing} at {Mid}-ocean {Ridges}}
  {Magma {Dynamics} with the {Enthalpy} {Method}: {Benchmark} {Solutions} and
  {Magmatic} {Focusing} at {Mid}-ocean {Ridges}}.{\BBCQ}
\newblock
\APACjournalVolNumPages{Journal of Petrology}{49}{12}{2099--2121}.
\newblock
\begin{APACrefURL}
  [{2020-01-10}]\url{https://academic.oup.com/petrology/article/49/12/2099/1531301}
  \end{APACrefURL}
\newblock
\begin{APACrefDOI} \doi{10.1093/petrology/egn058} \end{APACrefDOI}
\PrintBackRefs{\CurrentBib}

\bibitem [\protect \citeauthoryear {%
Keller%
, May%
\BCBL {}\ \BBA {} Kaus%
}{%
Keller%
\ \protect \BOthers {.}}{%
{\protect \APACyear {2013}}%
}]{%
keller_numerical_2013}
\APACinsertmetastar {%
keller_numerical_2013}%
\begin{APACrefauthors}%
Keller, T.%
, May, D\BPBI A.%
\BCBL {}\ \BBA {} Kaus, B\BPBI J\BPBI P.%
\end{APACrefauthors}%
\unskip\
\newblock
\APACrefYearMonthDay{2013}{{\APACmonth{12}}}{}.
\newblock
{\BBOQ}\APACrefatitle {Numerical modelling of magma dynamics coupled to
  tectonic deformation of lithosphere and crust} {Numerical modelling of magma
  dynamics coupled to tectonic deformation of lithosphere and crust}.{\BBCQ}
\newblock
\APACjournalVolNumPages{Geophysical Journal International}{195}{3}{1406--1442}.
\newblock
\begin{APACrefURL}
  [{2020-03-06}]\url{https://academic.oup.com/gji/article/195/3/1406/2874184}
  \end{APACrefURL}
\newblock
\APACrefnote{Publisher: Oxford Academic}
\newblock
\begin{APACrefDOI} \doi{10.1093/gji/ggt306} \end{APACrefDOI}
\PrintBackRefs{\CurrentBib}

\bibitem [\protect \citeauthoryear {%
Keszthelyi%
, Jaeger%
, Turtle%
, Milazzo%
\BCBL {}\ \BBA {} Radebaugh%
}{%
Keszthelyi%
\ \protect \BOthers {.}}{%
{\protect \APACyear {2004}}%
}]{%
keszthelyi_post-galileo_2004}
\APACinsertmetastar {%
keszthelyi_post-galileo_2004}%
\begin{APACrefauthors}%
Keszthelyi, L.%
, Jaeger, W\BPBI L.%
, Turtle, E\BPBI P.%
, Milazzo, M.%
\BCBL {}\ \BBA {} Radebaugh, J.%
\end{APACrefauthors}%
\unskip\
\newblock
\APACrefYearMonthDay{2004}{{\APACmonth{05}}}{}.
\newblock
{\BBOQ}\APACrefatitle {A post-{Galileo} view of {Io}'s interior} {A
  post-{Galileo} view of {Io}'s interior}.{\BBCQ}
\newblock
\APACjournalVolNumPages{Icarus}{169}{1}{271--286}.
\newblock
\begin{APACrefURL}
  [{2020-04-21}]\url{http://www.sciencedirect.com/science/article/pii/S0019103504000351}
  \end{APACrefURL}
\newblock
\begin{APACrefDOI} \doi{10.1016/j.icarus.2004.01.005} \end{APACrefDOI}
\PrintBackRefs{\CurrentBib}

\bibitem [\protect \citeauthoryear {%
Keszthelyi%
\ \BBA {} McEwen%
}{%
Keszthelyi%
\ \BBA {} McEwen%
}{%
{\protect \APACyear {1997}}%
}]{%
keszthelyi_magmatic_1997}
\APACinsertmetastar {%
keszthelyi_magmatic_1997}%
\begin{APACrefauthors}%
Keszthelyi, L.%
\BCBT {}\ \BBA {} McEwen, A.%
\end{APACrefauthors}%
\unskip\
\newblock
\APACrefYearMonthDay{1997}{{\APACmonth{12}}}{}.
\newblock
{\BBOQ}\APACrefatitle {Magmatic {Differentiation} of {Io}} {Magmatic
  {Differentiation} of {Io}}.{\BBCQ}
\newblock
\APACjournalVolNumPages{Icarus}{130}{2}{437--448}.
\newblock
\begin{APACrefURL}
  \url{http://www.sciencedirect.com/science/article/pii/S0019103597958371}
  \end{APACrefURL}
\newblock
\begin{APACrefDOI} \doi{10.1006/icar.1997.5837} \end{APACrefDOI}
\PrintBackRefs{\CurrentBib}

\bibitem [\protect \citeauthoryear {%
Khurana%
\ \protect \BOthers {.}}{%
Khurana%
\ \protect \BOthers {.}}{%
{\protect \APACyear {2011}}%
}]{%
khurana_evidence_2011}
\APACinsertmetastar {%
khurana_evidence_2011}%
\begin{APACrefauthors}%
Khurana, K\BPBI K.%
, Jia, X.%
, Kivelson, M\BPBI G.%
, Nimmo, F.%
, Schubert, G.%
\BCBL {}\ \BBA {} Russell, C\BPBI T.%
\end{APACrefauthors}%
\unskip\
\newblock
\APACrefYearMonthDay{2011}{{\APACmonth{06}}}{}.
\newblock
{\BBOQ}\APACrefatitle {Evidence of a {Global} {Magma} {Ocean} in {Io}'s
  {Interior}} {Evidence of a {Global} {Magma} {Ocean} in {Io}'s
  {Interior}}.{\BBCQ}
\newblock
\APACjournalVolNumPages{Science}{332}{6034}{1186--1189}.
\newblock
\begin{APACrefURL}
  [{2017-10-11}]\url{http://science.sciencemag.org/content/332/6034/1186}
  \end{APACrefURL}
\newblock
\begin{APACrefDOI} \doi{10.1126/science.1201425} \end{APACrefDOI}
\PrintBackRefs{\CurrentBib}

\bibitem [\protect \citeauthoryear {%
Kirchoff%
\ \BBA {} McKinnon%
}{%
Kirchoff%
\ \BBA {} McKinnon%
}{%
{\protect \APACyear {2009}}%
}]{%
kirchoff_formation_2009}
\APACinsertmetastar {%
kirchoff_formation_2009}%
\begin{APACrefauthors}%
Kirchoff, M\BPBI R.%
\BCBT {}\ \BBA {} McKinnon, W\BPBI B.%
\end{APACrefauthors}%
\unskip\
\newblock
\APACrefYearMonthDay{2009}{{\APACmonth{06}}}{}.
\newblock
{\BBOQ}\APACrefatitle {Formation of mountains on {Io}: {Variable} volcanism and
  thermal stresses} {Formation of mountains on {Io}: {Variable} volcanism and
  thermal stresses}.{\BBCQ}
\newblock
\APACjournalVolNumPages{Icarus}{201}{2}{598--614}.
\newblock
\begin{APACrefURL}
  \url{http://www.sciencedirect.com/science/article/pii/S0019103509000633}
  \end{APACrefURL}
\newblock
\begin{APACrefDOI} \doi{10.1016/j.icarus.2009.02.006} \end{APACrefDOI}
\PrintBackRefs{\CurrentBib}

\bibitem [\protect \citeauthoryear {%
Kirchoff%
, McKinnon%
\BCBL {}\ \BBA {} Bland%
}{%
Kirchoff%
\ \protect \BOthers {.}}{%
{\protect \APACyear {2020}}%
}]{%
kirchoff_effects_2020}
\APACinsertmetastar {%
kirchoff_effects_2020}%
\begin{APACrefauthors}%
Kirchoff, M\BPBI R.%
, McKinnon, W\BPBI B.%
\BCBL {}\ \BBA {} Bland, M\BPBI T.%
\end{APACrefauthors}%
\unskip\
\newblock
\APACrefYearMonthDay{2020}{{\APACmonth{01}}}{}.
\newblock
{\BBOQ}\APACrefatitle {Effects of faulting on crustal stresses during mountain
  formation on {Io}} {Effects of faulting on crustal stresses during mountain
  formation on {Io}}.{\BBCQ}
\newblock
\APACjournalVolNumPages{Icarus}{335}{}{113326}.
\newblock
\begin{APACrefURL}
  [{2020-02-27}]\url{http://www.sciencedirect.com/science/article/pii/S0019103518304020}
  \end{APACrefURL}
\newblock
\begin{APACrefDOI} \doi{10.1016/j.icarus.2019.05.028} \end{APACrefDOI}
\PrintBackRefs{\CurrentBib}

\bibitem [\protect \citeauthoryear {%
Kirchoff%
, McKinnon%
\BCBL {}\ \BBA {} Schenk%
}{%
Kirchoff%
\ \protect \BOthers {.}}{%
{\protect \APACyear {2011}}%
}]{%
kirchoff_global_2011}
\APACinsertmetastar {%
kirchoff_global_2011}%
\begin{APACrefauthors}%
Kirchoff, M\BPBI R.%
, McKinnon, W\BPBI B.%
\BCBL {}\ \BBA {} Schenk, P\BPBI M.%
\end{APACrefauthors}%
\unskip\
\newblock
\APACrefYearMonthDay{2011}{{\APACmonth{01}}}{}.
\newblock
{\BBOQ}\APACrefatitle {Global distribution of volcanic centers and mountains on
  {Io}: {Control} by asthenospheric heating and implications for mountain
  formation} {Global distribution of volcanic centers and mountains on {Io}:
  {Control} by asthenospheric heating and implications for mountain
  formation}.{\BBCQ}
\newblock
\APACjournalVolNumPages{Earth and Planetary Science Letters}{301}{1}{22--30}.
\newblock
\begin{APACrefURL}
  \url{http://www.sciencedirect.com/science/article/pii/S0012821X10007132}
  \end{APACrefURL}
\newblock
\begin{APACrefDOI} \doi{10.1016/j.epsl.2010.11.018} \end{APACrefDOI}
\PrintBackRefs{\CurrentBib}

\bibitem [\protect \citeauthoryear {%
Lainey%
, Arlot%
, Karatekin%
\BCBL {}\ \BBA {} Van~Hoolst%
}{%
Lainey%
\ \protect \BOthers {.}}{%
{\protect \APACyear {2009}}%
}]{%
lainey_strong_2009}
\APACinsertmetastar {%
lainey_strong_2009}%
\begin{APACrefauthors}%
Lainey, V.%
, Arlot, J\BHBI E.%
, Karatekin, Ã.%
\BCBL {}\ \BBA {} Van~Hoolst, T.%
\end{APACrefauthors}%
\unskip\
\newblock
\APACrefYearMonthDay{2009}{{\APACmonth{06}}}{}.
\newblock
{\BBOQ}\APACrefatitle {Strong tidal dissipation in {Io} and {Jupiter} from
  astrometric observations} {Strong tidal dissipation in {Io} and {Jupiter}
  from astrometric observations}.{\BBCQ}
\newblock
\APACjournalVolNumPages{Nature}{459}{7249}{957--959}.
\newblock
\begin{APACrefURL}
  [{2017-08-14}]\url{http://www.nature.com/nature/journal/v459/n7249/full/nature08108.html?foxtrotcallback=true}
  \end{APACrefURL}
\newblock
\begin{APACrefDOI} \doi{10.1038/nature08108} \end{APACrefDOI}
\PrintBackRefs{\CurrentBib}

\bibitem [\protect \citeauthoryear {%
Lichtenberg%
, Keller%
, Katz%
, Golabek%
\BCBL {}\ \BBA {} Gerya%
}{%
Lichtenberg%
\ \protect \BOthers {.}}{%
{\protect \APACyear {2019}}%
}]{%
lichtenberg_magma_2019}
\APACinsertmetastar {%
lichtenberg_magma_2019}%
\begin{APACrefauthors}%
Lichtenberg, T.%
, Keller, T.%
, Katz, R\BPBI F.%
, Golabek, G\BPBI J.%
\BCBL {}\ \BBA {} Gerya, T\BPBI V.%
\end{APACrefauthors}%
\unskip\
\newblock
\APACrefYearMonthDay{2019}{{\APACmonth{02}}}{}.
\newblock
{\BBOQ}\APACrefatitle {Magma ascent in planetesimals: {Control} by grain size}
  {Magma ascent in planetesimals: {Control} by grain size}.{\BBCQ}
\newblock
\APACjournalVolNumPages{Earth and Planetary Science Letters}{507}{}{154--165}.
\newblock
\begin{APACrefURL}
  [{2019-06-25}]\url{http://www.sciencedirect.com/science/article/pii/S0012821X18306939}
  \end{APACrefURL}
\newblock
\begin{APACrefDOI} \doi{10.1016/j.epsl.2018.11.034} \end{APACrefDOI}
\PrintBackRefs{\CurrentBib}

\bibitem [\protect \citeauthoryear {%
McKenzie%
}{%
McKenzie%
}{%
{\protect \APACyear {1984}}%
}]{%
mckenzie_generation_1984}
\APACinsertmetastar {%
mckenzie_generation_1984}%
\begin{APACrefauthors}%
McKenzie, D.%
\end{APACrefauthors}%
\unskip\
\newblock
\APACrefYearMonthDay{1984}{{\APACmonth{08}}}{}.
\newblock
{\BBOQ}\APACrefatitle {The {Generation} and {Compaction} of {Partially}
  {Molten} {Rock}} {The {Generation} and {Compaction} of {Partially} {Molten}
  {Rock}}.{\BBCQ}
\newblock
\APACjournalVolNumPages{Journal of Petrology}{25}{3}{713--765}.
\newblock
\begin{APACrefURL}
  [{2018-11-01}]\url{https://academic.oup.com/petrology/article/25/3/713/1394279}
  \end{APACrefURL}
\newblock
\begin{APACrefDOI} \doi{10.1093/petrology/25.3.713} \end{APACrefDOI}
\PrintBackRefs{\CurrentBib}

\bibitem [\protect \citeauthoryear {%
McKinnon%
, Schenk%
\BCBL {}\ \BBA {} Dombard%
}{%
McKinnon%
\ \protect \BOthers {.}}{%
{\protect \APACyear {2001}}%
}]{%
mckinnon_chaos_2001}
\APACinsertmetastar {%
mckinnon_chaos_2001}%
\begin{APACrefauthors}%
McKinnon, W\BPBI B.%
, Schenk, P\BPBI M.%
\BCBL {}\ \BBA {} Dombard, A\BPBI J.%
\end{APACrefauthors}%
\unskip\
\newblock
\APACrefYearMonthDay{2001}{{\APACmonth{02}}}{}.
\newblock
{\BBOQ}\APACrefatitle {Chaos on {Io}: {A} model for formation of mountain
  blocks by crustal heating, melting, and tilting} {Chaos on {Io}: {A} model
  for formation of mountain blocks by crustal heating, melting, and
  tilting}.{\BBCQ}
\newblock
\APACjournalVolNumPages{Geology}{29}{2}{103--106}.
\newblock
\begin{APACrefURL}
  [{2019-05-24}]\url{https://pubs.geoscienceworld.org/gsa/geology/article/29/2/103/191954/chaos-on-io-a-model-for-formation-of-mountain}
  \end{APACrefURL}
\newblock
\begin{APACrefDOI} \doi{10.1130/0091-7613(2001)029<0103:COIAMF>2.0.CO;2}
  \end{APACrefDOI}
\PrintBackRefs{\CurrentBib}

\bibitem [\protect \citeauthoryear {%
Moore%
}{%
Moore%
}{%
{\protect \APACyear {2001}}%
}]{%
moore_thermal_2001}
\APACinsertmetastar {%
moore_thermal_2001}%
\begin{APACrefauthors}%
Moore, W\BPBI B.%
\end{APACrefauthors}%
\unskip\
\newblock
\APACrefYearMonthDay{2001}{{\APACmonth{12}}}{}.
\newblock
{\BBOQ}\APACrefatitle {The {Thermal} {State} of {Io}} {The {Thermal} {State} of
  {Io}}.{\BBCQ}
\newblock
\APACjournalVolNumPages{Icarus}{154}{2}{548--550}.
\newblock
\begin{APACrefURL}
  \url{http://www.sciencedirect.com/science/article/pii/S0019103501967399}
  \end{APACrefURL}
\newblock
\begin{APACrefDOI} \doi{10.1006/icar.2001.6739} \end{APACrefDOI}
\PrintBackRefs{\CurrentBib}

\bibitem [\protect \citeauthoryear {%
Moore%
}{%
Moore%
}{%
{\protect \APACyear {2003}}%
}]{%
moore_tidal_2003}
\APACinsertmetastar {%
moore_tidal_2003}%
\begin{APACrefauthors}%
Moore, W\BPBI B.%
\end{APACrefauthors}%
\unskip\
\newblock
\APACrefYearMonthDay{2003}{{\APACmonth{08}}}{}.
\newblock
{\BBOQ}\APACrefatitle {Tidal heating and convection in {Io}} {Tidal heating and
  convection in {Io}}.{\BBCQ}
\newblock
\APACjournalVolNumPages{Journal of Geophysical Research:
  Planets}{108}{E8}{5096}.
\newblock
\begin{APACrefURL}
  \url{http://onlinelibrary.wiley.com/doi/10.1029/2002JE001943/abstract}
  \end{APACrefURL}
\newblock
\begin{APACrefDOI} \doi{10.1029/2002JE001943} \end{APACrefDOI}
\PrintBackRefs{\CurrentBib}

\bibitem [\protect \citeauthoryear {%
O'Reilly%
\ \BBA {} Davies%
}{%
O'Reilly%
\ \BBA {} Davies%
}{%
{\protect \APACyear {1981}}%
}]{%
oreilly_magma_1981}
\APACinsertmetastar {%
oreilly_magma_1981}%
\begin{APACrefauthors}%
O'Reilly, T\BPBI C.%
\BCBT {}\ \BBA {} Davies, G\BPBI F.%
\end{APACrefauthors}%
\unskip\
\newblock
\APACrefYearMonthDay{1981}{{\APACmonth{04}}}{}.
\newblock
{\BBOQ}\APACrefatitle {Magma transport of heat on {Io}: {A} mechanism allowing
  a thick lithosphere} {Magma transport of heat on {Io}: {A} mechanism allowing
  a thick lithosphere}.{\BBCQ}
\newblock
\APACjournalVolNumPages{Geophysical Research Letters}{8}{4}{313--316}.
\newblock
\begin{APACrefURL}
  \url{http://onlinelibrary.wiley.com/doi/10.1029/GL008i004p00313/abstract}
  \end{APACrefURL}
\newblock
\begin{APACrefDOI} \doi{10.1029/GL008i004p00313} \end{APACrefDOI}
\PrintBackRefs{\CurrentBib}

\bibitem [\protect \citeauthoryear {%
Rathbun%
, Lopes%
\BCBL {}\ \BBA {} Spencer%
}{%
Rathbun%
\ \protect \BOthers {.}}{%
{\protect \APACyear {2018}}%
}]{%
rathbun_global_2018}
\APACinsertmetastar {%
rathbun_global_2018}%
\begin{APACrefauthors}%
Rathbun, J\BPBI A.%
, Lopes, R\BPBI M\BPBI C.%
\BCBL {}\ \BBA {} Spencer, J\BPBI R.%
\end{APACrefauthors}%
\unskip\
\newblock
\APACrefYearMonthDay{2018}{{\APACmonth{10}}}{}.
\newblock
{\BBOQ}\APACrefatitle {The {Global} {Distribution} of {Active} {Ionian}
  {Volcanoes} and {Implications} for {Tidal} {Heating} {Models}} {The {Global}
  {Distribution} of {Active} {Ionian} {Volcanoes} and {Implications} for
  {Tidal} {Heating} {Models}}.{\BBCQ}
\newblock
\APACjournalVolNumPages{The Astronomical Journal}{156}{5}{207}.
\newblock
\begin{APACrefURL}
  [{2020-04-22}]\url{https://doi.org/10.3847%2F1538-3881%2Faae370}
  \end{APACrefURL}
\newblock
\APACrefnote{Publisher: American Astronomical Society}
\newblock
\begin{APACrefDOI} \doi{10.3847/1538-3881/aae370} \end{APACrefDOI}
\PrintBackRefs{\CurrentBib}

\bibitem [\protect \citeauthoryear {%
Renaud%
\ \BBA {} Henning%
}{%
Renaud%
\ \BBA {} Henning%
}{%
{\protect \APACyear {2018}}%
}]{%
renaud_increased_2018}
\APACinsertmetastar {%
renaud_increased_2018}%
\begin{APACrefauthors}%
Renaud, J\BPBI P.%
\BCBT {}\ \BBA {} Henning, W\BPBI G.%
\end{APACrefauthors}%
\unskip\
\newblock
\APACrefYearMonthDay{2018}{{\APACmonth{04}}}{}.
\newblock
{\BBOQ}\APACrefatitle {Increased {Tidal} {Dissipation} {Using} {Advanced}
  {Rheological} {Models}: {Implications} for {Io} and {Tidally} {Active}
  {Exoplanets}} {Increased {Tidal} {Dissipation} {Using} {Advanced}
  {Rheological} {Models}: {Implications} for {Io} and {Tidally} {Active}
  {Exoplanets}}.{\BBCQ}
\newblock
\APACjournalVolNumPages{The Astrophysical Journal}{857}{2}{98}.
\newblock
\begin{APACrefURL}
  [{2019-04-26}]\url{https://doi.org/10.3847%2F1538-4357%2Faab784}
  \end{APACrefURL}
\newblock
\begin{APACrefDOI} \doi{10.3847/1538-4357/aab784} \end{APACrefDOI}
\PrintBackRefs{\CurrentBib}

\bibitem [\protect \citeauthoryear {%
Rudge%
}{%
Rudge%
}{%
{\protect \APACyear {2018}}%
}]{%
rudge_viscosities_2018}
\APACinsertmetastar {%
rudge_viscosities_2018}%
\begin{APACrefauthors}%
Rudge, J\BPBI F.%
\end{APACrefauthors}%
\unskip\
\newblock
\APACrefYearMonthDay{2018}{}{}.
\newblock
{\BBOQ}\APACrefatitle {The {Viscosities} of {Partially} {Molten} {Materials}
  {Undergoing} {Diffusion} {Creep}} {The {Viscosities} of {Partially} {Molten}
  {Materials} {Undergoing} {Diffusion} {Creep}}.{\BBCQ}
\newblock
\APACjournalVolNumPages{Journal of Geophysical Research: Solid
  Earth}{123}{12}{10,534--10,562}.
\newblock
\begin{APACrefURL}
  [{2019-07-17}]\url{https://agupubs.onlinelibrary.wiley.com/doi/abs/10.1029/2018JB016530}
  \end{APACrefURL}
\newblock
\begin{APACrefDOI} \doi{10.1029/2018JB016530} \end{APACrefDOI}
\PrintBackRefs{\CurrentBib}

\bibitem [\protect \citeauthoryear {%
Schenk%
\ \BBA {} Bulmer%
}{%
Schenk%
\ \BBA {} Bulmer%
}{%
{\protect \APACyear {1998}}%
}]{%
schenk_origin_1998}
\APACinsertmetastar {%
schenk_origin_1998}%
\begin{APACrefauthors}%
Schenk, P\BPBI M.%
\BCBT {}\ \BBA {} Bulmer, M\BPBI H.%
\end{APACrefauthors}%
\unskip\
\newblock
\APACrefYearMonthDay{1998}{{\APACmonth{03}}}{}.
\newblock
{\BBOQ}\APACrefatitle {Origin of {Mountains} on {Io} by {Thrust} {Faulting} and
  {Large}-{Scale} {Mass} {Movements}} {Origin of {Mountains} on {Io} by
  {Thrust} {Faulting} and {Large}-{Scale} {Mass} {Movements}}.{\BBCQ}
\newblock
\APACjournalVolNumPages{Science}{279}{5356}{1514--1517}.
\newblock
\begin{APACrefURL}
  [{2017-10-11}]\url{http://science.sciencemag.org/content/279/5356/1514}
  \end{APACrefURL}
\newblock
\begin{APACrefDOI} \doi{10.1126/science.279.5356.1514} \end{APACrefDOI}
\PrintBackRefs{\CurrentBib}

\bibitem [\protect \citeauthoryear {%
Sparks%
\ \BBA {} Parmentier%
}{%
Sparks%
\ \BBA {} Parmentier%
}{%
{\protect \APACyear {1991}}%
}]{%
sparks_melt_1991}
\APACinsertmetastar {%
sparks_melt_1991}%
\begin{APACrefauthors}%
Sparks, D\BPBI W.%
\BCBT {}\ \BBA {} Parmentier, E\BPBI M.%
\end{APACrefauthors}%
\unskip\
\newblock
\APACrefYearMonthDay{1991}{{\APACmonth{08}}}{}.
\newblock
{\BBOQ}\APACrefatitle {Melt extraction from the mantle beneath spreading
  centers} {Melt extraction from the mantle beneath spreading centers}.{\BBCQ}
\newblock
\APACjournalVolNumPages{Earth and Planetary Science Letters}{105}{4}{368--377}.
\newblock
\begin{APACrefURL}
  [{2019-06-21}]\url{http://www.sciencedirect.com/science/article/pii/0012821X9190178K}
  \end{APACrefURL}
\newblock
\begin{APACrefDOI} \doi{10.1016/0012-821X(91)90178-K} \end{APACrefDOI}
\PrintBackRefs{\CurrentBib}

\bibitem [\protect \citeauthoryear {%
Spencer%
, Katz%
\BCBL {}\ \BBA {} Hewitt%
}{%
Spencer%
\ \protect \BOthers {.}}{%
{\protect \APACyear {2020}}%
}]{%
spencer_spencer-spaceio_compaction_2020}
\APACinsertmetastar {%
spencer_spencer-spaceio_compaction_2020}%
\begin{APACrefauthors}%
Spencer, D\BPBI C.%
, Katz, R\BPBI F.%
\BCBL {}\ \BBA {} Hewitt, I\BPBI J.%
\end{APACrefauthors}%
\unskip\
\newblock
\APACrefYearMonthDay{2020}{{\APACmonth{04}}}{}.
\newblock
\APACrefbtitle {Spencer-space/io\_compaction: io\_compaction v1.1.0.}
  {Spencer-space/io\_compaction: io\_compaction v1.1.0.}
\newblock
\APACaddressPublisher{}{Zenodo}.
\newblock
\begin{APACrefURL} [{2020-04-27}]\url{https://zenodo.org/record/3769022}
  \end{APACrefURL}
\newblock
\begin{APACrefDOI} \doi{10.5281/zenodo.3769022} \end{APACrefDOI}
\PrintBackRefs{\CurrentBib}

\bibitem [\protect \citeauthoryear {%
Spiegelman%
}{%
Spiegelman%
}{%
{\protect \APACyear {1993}}%
}]{%
spiegelman_flow_1993}
\APACinsertmetastar {%
spiegelman_flow_1993}%
\begin{APACrefauthors}%
Spiegelman, M.%
\end{APACrefauthors}%
\unskip\
\newblock
\APACrefYearMonthDay{1993}{{\APACmonth{02}}}{}.
\newblock
{\BBOQ}\APACrefatitle {Flow in deformable porous media. {Part} 2 {Numerical}
  analysis – the relationship between shock waves and solitary waves} {Flow
  in deformable porous media. {Part} 2 {Numerical} analysis – the
  relationship between shock waves and solitary waves}.{\BBCQ}
\newblock
\APACjournalVolNumPages{Journal of Fluid Mechanics}{247}{}{39--63}.
\newblock
\begin{APACrefURL}
  [{2020-02-27}]\url{https://www.cambridge.org/core/journals/journal-of-fluid-mechanics/article/flow-in-deformable-porous-media-part-2-numerical-analysis-the-relationship-between-shock-waves-and-solitary-waves/A4AAC407AE3B1016AED122EBE70D50C0}
  \end{APACrefURL}
\newblock
\APACrefnote{Publisher: Cambridge University Press}
\newblock
\begin{APACrefDOI} \doi{10.1017/S0022112093000370} \end{APACrefDOI}
\PrintBackRefs{\CurrentBib}

\bibitem [\protect \citeauthoryear {%
Steinke%
, Hu%
, Höning%
, van~der Wal%
\BCBL {}\ \BBA {} Vermeersen%
}{%
Steinke%
\ \protect \BOthers {.}}{%
{\protect \APACyear {2020}}%
}]{%
steinke_tidally_2020}
\APACinsertmetastar {%
steinke_tidally_2020}%
\begin{APACrefauthors}%
Steinke, T.%
, Hu, H.%
, Höning, D.%
, van~der Wal, W.%
\BCBL {}\ \BBA {} Vermeersen, B.%
\end{APACrefauthors}%
\unskip\
\newblock
\APACrefYearMonthDay{2020}{{\APACmonth{01}}}{}.
\newblock
{\BBOQ}\APACrefatitle {Tidally induced lateral variations of {Io}'s interior}
  {Tidally induced lateral variations of {Io}'s interior}.{\BBCQ}
\newblock
\APACjournalVolNumPages{Icarus}{335}{}{113299}.
\newblock
\begin{APACrefURL}
  [{2020-01-10}]\url{http://www.sciencedirect.com/science/article/pii/S0019103518307632}
  \end{APACrefURL}
\newblock
\begin{APACrefDOI} \doi{10.1016/j.icarus.2019.05.001} \end{APACrefDOI}
\PrintBackRefs{\CurrentBib}

\bibitem [\protect \citeauthoryear {%
Turner%
, Katz%
, Behn%
\BCBL {}\ \BBA {} Keller%
}{%
Turner%
\ \protect \BOthers {.}}{%
{\protect \APACyear {2017}}%
}]{%
turner_magmatic_2017}
\APACinsertmetastar {%
turner_magmatic_2017}%
\begin{APACrefauthors}%
Turner, A\BPBI J.%
, Katz, R\BPBI F.%
, Behn, M\BPBI D.%
\BCBL {}\ \BBA {} Keller, T.%
\end{APACrefauthors}%
\unskip\
\newblock
\APACrefYearMonthDay{2017}{}{}.
\newblock
{\BBOQ}\APACrefatitle {Magmatic {Focusing} to {Mid}-{Ocean} {Ridges}: {The}
  {Role} of {Grain}-{Size} {Variability} and {Non}-{Newtonian} {Viscosity}}
  {Magmatic {Focusing} to {Mid}-{Ocean} {Ridges}: {The} {Role} of
  {Grain}-{Size} {Variability} and {Non}-{Newtonian} {Viscosity}}.{\BBCQ}
\newblock
\APACjournalVolNumPages{Geochemistry, Geophysics,
  Geosystems}{18}{12}{4342--4355}.
\newblock
\begin{APACrefURL}
  [{2020-01-10}]\url{https://agupubs.onlinelibrary.wiley.com/doi/abs/10.1002/2017GC007048}
  \end{APACrefURL}
\newblock
\begin{APACrefDOI} \doi{10.1002/2017GC007048} \end{APACrefDOI}
\PrintBackRefs{\CurrentBib}

\bibitem [\protect \citeauthoryear {%
Tyler%
, Henning%
\BCBL {}\ \BBA {} Hamilton%
}{%
Tyler%
\ \protect \BOthers {.}}{%
{\protect \APACyear {2015}}%
}]{%
tyler_tidal_2015}
\APACinsertmetastar {%
tyler_tidal_2015}%
\begin{APACrefauthors}%
Tyler, R\BPBI H.%
, Henning, W\BPBI G.%
\BCBL {}\ \BBA {} Hamilton, C\BPBI W.%
\end{APACrefauthors}%
\unskip\
\newblock
\APACrefYearMonthDay{2015}{}{}.
\newblock
{\BBOQ}\APACrefatitle {Tidal {Heating} in a {Magma} {Ocean} within {Jupiter}'s
  {Moon} {Io}} {Tidal {Heating} in a {Magma} {Ocean} within {Jupiter}'s {Moon}
  {Io}}.{\BBCQ}
\newblock
\APACjournalVolNumPages{The Astrophysical Journal Supplement
  Series}{218}{2}{22}.
\newblock
\begin{APACrefURL} \url{http://stacks.iop.org/0067-0049/218/i=2/a=22}
  \end{APACrefURL}
\newblock
\begin{APACrefDOI} \doi{10.1088/0067-0049/218/2/22} \end{APACrefDOI}
\PrintBackRefs{\CurrentBib}

\bibitem [\protect \citeauthoryear {%
Veeder%
\ \protect \BOthers {.}}{%
Veeder%
\ \protect \BOthers {.}}{%
{\protect \APACyear {2012}}%
}]{%
veeder_io:_2012}
\APACinsertmetastar {%
veeder_io:_2012}%
\begin{APACrefauthors}%
Veeder, G\BPBI J.%
, Davies, A\BPBI G.%
, Matson, D\BPBI L.%
, Johnson, T\BPBI V.%
, Williams, D\BPBI A.%
\BCBL {}\ \BBA {} Radebaugh, J.%
\end{APACrefauthors}%
\unskip\
\newblock
\APACrefYearMonthDay{2012}{{\APACmonth{06}}}{}.
\newblock
{\BBOQ}\APACrefatitle {Io: {Volcanic} thermal sources and global heat flow}
  {Io: {Volcanic} thermal sources and global heat flow}.{\BBCQ}
\newblock
\APACjournalVolNumPages{Icarus}{219}{2}{701--722}.
\newblock
\begin{APACrefURL}
  \url{http://www.sciencedirect.com/science/article/pii/S0019103512001339}
  \end{APACrefURL}
\newblock
\begin{APACrefDOI} \doi{10.1016/j.icarus.2012.04.004} \end{APACrefDOI}
\PrintBackRefs{\CurrentBib}

\bibitem [\protect \citeauthoryear {%
Williams%
\ \protect \BOthers {.}}{%
Williams%
\ \protect \BOthers {.}}{%
{\protect \APACyear {2011}}%
}]{%
williams_volcanism_2011}
\APACinsertmetastar {%
williams_volcanism_2011}%
\begin{APACrefauthors}%
Williams, D\BPBI A.%
, Keszthelyi, L\BPBI P.%
, Crown, D\BPBI A.%
, Yff, J\BPBI A.%
, Jaeger, W\BPBI L.%
, Schenk, P\BPBI M.%
\BDBL {}Becker, T\BPBI L.%
\end{APACrefauthors}%
\unskip\
\newblock
\APACrefYearMonthDay{2011}{{\APACmonth{07}}}{}.
\newblock
{\BBOQ}\APACrefatitle {Volcanism on {Io}: {New} insights from global geologic
  mapping} {Volcanism on {Io}: {New} insights from global geologic
  mapping}.{\BBCQ}
\newblock
\APACjournalVolNumPages{Icarus}{214}{1}{91--112}.
\newblock
\begin{APACrefURL}
  [{2017-10-20}]\url{https://ezproxy-prd.bodleian.ox.ac.uk:2073/science/article/pii/S0019103511001710}
  \end{APACrefURL}
\newblock
\begin{APACrefDOI} \doi{10.1016/j.icarus.2011.05.007} \end{APACrefDOI}
\PrintBackRefs{\CurrentBib}

\end{thebibliography}

%
%
%
%
%

\end{document}